\pdfoutput=1
\documentclass[12pt]{article}

\usepackage{a4}
\usepackage{amsfonts,amssymb,amsmath,amsthm}
\usepackage[colorlinks=true,linkcolor=black,citecolor=teal,urlcolor=blue,filecolor=black]{hyperref}
\usepackage{authblk}
\usepackage{graphicx,subcaption,tikz,pgfplots}
\pgfplotsset{compat=1.16}
\usepackage{xcolor,multirow}
\usepackage[section]{placeins}
\usepackage[version=4]{mhchem}

\newcommand{\refcite}[1]{\cite{#1}}
\newcommand{\tbl}[2]{\caption{#1}#2}
\newcommand{\startappendix}{\appendix\section*{APPENDICES}}
\newcommand{\myappendix}[1]{\section{#1}}

\begin{document}
\title{Carbon-made Lobachevsky surfaces}

\author[1]{R. Gabbrielli}
\author[1,2]{P. Pais}
\author[1,3,4]{P. Castorina}
\author[5]{T. P. C. Klaver}
\author[5,6]{D. Legut}
\author[7]{O. De Luca}
\author[7]{R. G. Agostino}
\author[8,1,4]{A. Iorio}

\affil[1]{Institute of Particle and Nuclear Physics, Faculty of Mathematics and Physics, Charles University, V Hole\v{s}ovi\v{c}k\'{a}ch 2, 18000 Prague 8, Czech Republic}
\affil[2]{Instituto de Ciencias Físicas y Matemáticas, Universidad Austral de Chile, Casilla 567, 5090000 Valdivia, Chile}
\affil[3]{Istituto Nazionale di Fisica Nucleare, Sezione di Catania, Via Santa Sofia 64, 95123 Catania, Italy}
\affil[4]{Democritos Technologies, Rybn\'a 24, 11000 Prague 1, Czech Republic}
\affil[5]{IT4Innovations, VSB - Technical University of Ostrava, 17. listopadu 2172/15, 708 00 Ostrava, Czech Republic}
\affil[6]{Department of Condensed Matter Physics, Faculty of Mathematics and Physics, Charles University, Ke Karlovu 3, 12116 Prague 2, Czech Republic}
\affil[7]{Department of Physics and STAR Research Infrastructure, University of Calabria, 87036 Rende (CS), Italy}
\affil[8]{Istituto Nazionale di Fisica Nucleare, Sezione di Bologna, Viale Berti Pichat 6/2, 40126 Bologna, Italy}
\maketitle

\begin{abstract}
Carbon-made surfaces of constant negative Gaussian curvature, $K<0$, realize portions of the Lobachevsky plane in the laboratory. The motivations range from analogue gravity, where the three pseudospheres, elliptic (ELL), Beltrami (BEL) and hyperbolic (HYP), correspond to the three static phases of the spectrum of the Ba\~{n}ados--Teitelboim--Zanelli black hole, to water desalination, where the hourglass-shaped HYP is a natural candidate filter. We review and extend what is known about the construction of these structures, organizing the discussion around a single principle: the \textit{balanced Gauss--Bonnet theorem}. The total curvature fixed by the topology is shared among three contributions: the intrinsic curvature carried by the disclination defects (heptagons, octagons, \dots), the ``non-topological'' intrinsic curvature carried by elastically deformed hexagons, and the boundary terms. We show that curvature without defects cannot change the topology: a sphere made of hexagons only is a punctured sphere, as we prove with an explicit polyhedral construction whose deficit angles add up to $4\pi$, ring after ring of hexagons; and we show that defects take over above a threshold that is essentially geometric. We then show that the discrete symmetries of the defects on the three pseudospheres stem from the elliptic, parabolic and hyperbolic one-parameter subgroups of $SL(2,\mathbb{R})$: we construct them for the family of ELLs, $F_{C_{k-1}}$, recover the known group of BEL, $F_Y$, as their limit, and propose the one for HYP, $F_{D_6}$. Finally, we confront these ideal arrangements with carbon structures obtained by design, as polycyclic aromatic hydrocarbons, and by simulated nano-extrusion and annealing, where the same total topological charge is distributed in very different ways, and the desired shape is obtained only when the boundary term is under control. We close by suggesting that, while the finite subgroups of $SO(3)$ are tied to the five Platonic solids, the infinitely many discrete (Fuchsian) subgroups of $SO(2,1)$ may be in correspondence with the infinitely many Lobachevsky surfaces, of which the three pseudospheres are the first instance.
\end{abstract}

\tableofcontents

\section{Introduction \label{sec:intro}}

\begin{figure}[htbp]
  \centering
  \includegraphics[width=0.7\linewidth]{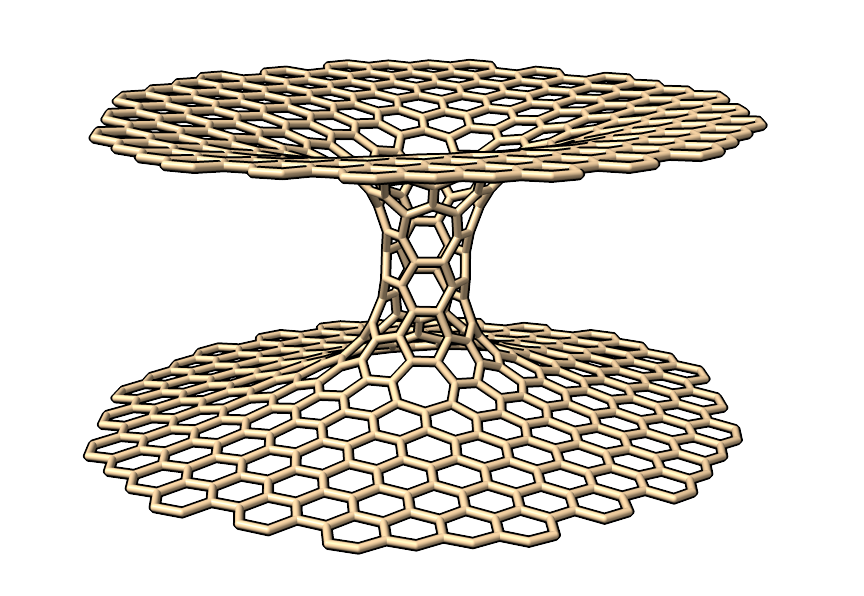}
  \caption{A carbon-made hyperbolic pseudosphere (HYP) joining two flat graphene sheets. The structure contains exactly twelve heptagons, i.e., the twelve units of curvature $-\pi/3$ that the balanced Gauss--Bonnet theorem requires for HYP, $-4\pi$ in total (Sections~\ref{sec:PseudoLimits} and~\ref{sec:HYP}), while all the other rings are hexagons. The same geometry is a natural candidate for desalination filters (Section~\ref{sec:Desalination}).}
  \label{fig:intro}
\end{figure}

The main message of this work is that \textit{a surface is defined primarily by its topology, and only to a minor extent by its local geometry}, provided that ``topology'' is understood in the precise sense of the Gauss--Bonnet theorem, boundaries included. When the surface is to be built from a discrete material, such as graphene, this statement becomes very concrete: the total curvature required by the topology must be carried by something, and that something can be disclination defects, elastically deformed hexagons, or the boundaries. Which of the three carries what, and how, decides whether the structure we obtain is the one we wanted. We shall call this the \textit{balanced} Gauss--Bonnet theorem, and we shall use it as the guiding principle throughout. A carbon structure that realizes this idea is shown in Fig.~\ref{fig:intro}.

Our main theoretical motivation is the use of graphene as an analogue of gravitational phenomena, both classical and quantum \cite{weylgraphene,iorio2015,universegraphene}. This includes the Hawking effect, noncommutativity, generalized uncertainty principles (GUPs), teleparallel gravity and many other phenomena \cite{i2,iorio2012,iorio2014,iorio2018generalized,iorio2018anti,ip5,iorio2020generalized,ciappina2020torsion,threelayers,NCthreelayers}. Those results are based on the fact that graphene low-energy quasiparticles are massless Dirac pseudo-relativistic fermions~\cite{PacoReview2009} (the matter field $\psi$), propagating on a carbon two-dimensional honeycomb lattice that, for large electronic wavelengths, well approximate the (spatial part of) continuous three-dimensional spacetimes~\cite{i2} (the metric $g_{\mu \nu}$). The combined system, $(\psi, g_{\mu \nu})$, realizes in the lab a quantum field on a curved classical background and, when higher-energy lattice effects are considered, quantum gravity scenarios too.

In particular, we are motivated by the recent theoretical results of Ref.~\protect\refcite{iorio2026classical}, which present a novel approach (compared to the one that uses Weyl symmetry \cite{weylgraphene}, employed earlier, see, e.g., the reviews \cite{iorio2015,universegraphene}) to extract information on gravitational physics from suitably shaped graphene membranes. This approach emphasizes the role of the three pseudospheres, the elliptic (ELL), hyperbolic (HYP) and Beltrami (BEL) pseudospheres, in relation to three (static and spinless) phases of the dynamics of the one physical system, that is the three-dimensional black hole of Ba\~{n}ados, Teitelboim and Zanelli (BTZ) \cite{BTZ1992}. A prominent role there is played by HYP \cite{iorio2026classical}. Indeed, in Ref.~\protect\refcite{klaver2025stability} the first carbon-made HYP was realized in a simulation lab and studied with the aid of a supercomputer.

As ELL, HYP and BEL are three important representatives of the infinite family of ``Lobachevsky surfaces'' \cite{ovchinnikov,mclachlan} (that is, the surfaces of constant negative Gaussian curvature, $K$), a general and mathematical motivation of this study is to establish a link between defects structure, arrangement and symmetry group and specific surfaces. In turn, given the correspondence between such surfaces and solitons \cite{mclachlan}, theoretical motivations lie in that area as well.

Motivations rooted in gravitational physics, nonlinear systems or pure mathematics are not the only ones. The management of the geometry and topology of carbon structures is important in a variety of other contexts too, both theoretical and practical. One example, of vital practical importance, is carbon desalination filters. Indeed, among the innumerable applications of graphene and of other allotropic forms of carbon, such as nanotubes, nanocones and others, their use as a ``molecular sieve'' for filtering polluted, fresh or marine waters, to obtain clean or drinkable water, is by now an established reality, as reported in a large literature; see, e.g., Ref.~\protect\refcite{GeimScience2014} and Ref.~\protect\refcite{Castelletto2021}. To this end, the geometry of such carbon structures plays a very important role; a particularly interesting case is the \textit{hourglass-shaped} carbon structures \cite{hourglass2021,hourglass2025}. The carbon-made HYP, with its hourglass shape, is a natural candidate for a serious study of the role of defects in the desalination process \cite{Iorio2024DesalinationPseudospheres}.

The paper is organized as follows. In Section~\ref{sec:TopologicalSurfaces} we recall the key facts of the topology of surfaces and of their tessellations, from the Platonic solids to the hexagonal lattice. In Section~\ref{sec:DefectFreeCurvature} we discuss intrinsic curvature without topological defects, and show that it cannot change the topology: a sphere made of hexagons only is a punctured sphere. In Section~\ref{sec:LobachevksyPlatonicSolids} we introduce the Lobachevsky geometry, its symmetry group $SL(2,\mathbb{R})\sim SO(2,1)$, and its discrete subgroups, and we show how the elliptic, parabolic and hyperbolic one-parameter subgroups single out the three pseudospheres, whose geometry and Gauss--Bonnet balance are discussed in Section~\ref{PseudospheresMathsGeneral}. Sections~\ref{sec:ELL}, \ref{sec:BEL} and \ref{sec:HYP} are devoted to carbon-made ELL, BEL and HYP, respectively: the ideal arrangements of defects and their discrete symmetries, and the actual carbon structures obtained by design or by simulated growth. In Section~\ref{sec:Applications} we outline the use of these structures for analogue gravity (the BTZ spectrum and wormholes) and for desalination. Our conclusions are in Section~\ref{sec:Summary}. Two appendices collect the deficit-angle formulae for discrete structures and an explicit counting of the curvature of a simulated carbon HYP.

\section{Topology of surfaces} \label{sec:TopologicalSurfaces}

Let us set up the stage, recalling key facts of elementary topology\footnote{We take the view that ``elementary'' means fundamental, not trivial.}. In particular, we shall discuss why \textit{a surface is defined primarily by its topology} and only to a minor extent by its local geometry. An illustrative example, for instance, is the punctured sphere: it is topologically equivalent to a disk, not to a sphere, even though it is locally everywhere a sphere, except for a single point of zero measure. Details of this fascinating case are given in Section~\ref{sec:DefectFreeCurvature}.

This fact carries on to the discrete structures associated with that given surface, e.g., the Platonic solids corresponding to the sphere. They are themselves characterized by the type and spatial distribution of their topological defects, in turn descending from the topology of the surface of which they are the ``discrete'' version. For instance, to stay with the previous example, a dodecahedron with a missing pentagon is a possible discrete version of the punctured sphere, not of the sphere.

When it comes to discrete structures, we shall mostly discuss hexagonal lattices, because we have graphene as our focus, but the main results apply to the general case.

The five Platonic solids are the only regular polyhedra with the same topology of the sphere. Their Euler characteristic, $\chi$, is the same as that of the sphere
\begin{equation}\label{chiSpherechiPlao}
  \chi ({\rm sphere}) = 2 = \chi ({\rm Plato}) \,.
\end{equation}

For the sphere, one can see this from the Gauss--Bonnet theorem \cite{Eguchi:1980jx} for a manifold $\cal M$, that in its general form is
\begin{equation}\label{GaussBonnetFull}
  \int_{\cal M} K d^2 \mu + \int_{\partial \cal M} k_g d \ell = 2 \pi \, \chi (\cal M) \,,
\end{equation}
with $K$ the Gaussian curvature and $k_g$ the geodesic curvature of the curve $\partial \cal M$, that is the boundary of $\cal M$. The formula, when applied to a sphere (that has no boundary $\partial {\cal M}$) of constant radius $R$, gives simply the total curvature
\begin{equation}\label{GaussBonnetPartial}
K_{tot} ({\rm sphere}) =  \int_{{\rm sphere}} K d^2 \mu = \int_0^{2 \pi} d v \int_{- R \pi/2}^{+ R \pi/2} d u \, (R \cos u/R) \, \frac{1}{R^2}
  = 4 \pi \,,
\end{equation}
where $K = + 1/R^2$. Henceforth, being $K_{tot} = 2 \pi \chi$, we have
\begin{equation}\label{chiSphere}
 \chi ({\rm sphere}) = 2 \,.
\end{equation}

For the polyhedra one can see this from the Euler--Poincar\'e formula for the tiling of two-manifolds
\begin{equation}\label{EulPoinPolyhedra}
  V - E + F = 2 = \chi ({\rm Plato}) \,.
\end{equation}
It is easy to show that the five Platonic solids are the only regular polyhedra satisfying this formula.

As ELL, BEL and HYP have boundaries, henceforth the full theorem \eqref{GaussBonnetFull} applies, it is useful to recall that
\begin{equation}
  \chi ({\cal M}) = 2 - 2 g - b \,,
\end{equation}
with $g$ the genus and $b$ the number of boundaries of the orientable manifold $\cal M$. Examples are in Table~\ref{tab:VariousEuler}.

Two remarks on the boundary term of \eqref{GaussBonnetFull} are in order, as it plays a major role in what follows. First, if $\partial \cal M$ is only piecewise smooth, the boundary term also includes the exterior angles at its corners \cite{Eguchi:1980jx},
\begin{equation}\label{GaussBonnetCorners}
  \int_{\cal M} K d^2 \mu + \int_{\partial \cal M} k_g d \ell + \sum_i \left(\pi - \theta_i\right) = 2 \pi \, \chi (\cal M) \,,
\end{equation}
with $\theta_i$ the interior angle at the $i$th corner. For a flat polygon with straight sides, for instance, $K=0$ and $k_g=0$, and \eqref{GaussBonnetCorners} is the elementary statement that the exterior angles add up to $2\pi$, with $\chi=1$. Second, for a discrete structure, such as a carbon lattice, both terms become sums of angles: the bulk term is the sum of the deficit angles at the \textit{interior} vertices, $2\pi - \sum_i\theta_i$, that we shall use extensively from Section~\ref{sec:ThreeDeformations} on, while the boundary term is entirely of the kind of the last term of \eqref{GaussBonnetCorners}, i.e., it is the sum, over the \textit{boundary} vertices, of $\pi - \sum_i\theta_i$, where the $\theta_i$ are now the angles of the polygons meeting at that vertex (see Appendix~\ref{app:DeficitAngles}). The two are the discrete versions of $K$ and of $k_g$, respectively. As an illustration, in a flat flake of graphene the interior vertices carry no deficit angle, a boundary vertex belonging to one hexagon only contributes $\pi - \frac{2}{3}\pi = +\frac{\pi}{3}$, and a boundary vertex shared by two hexagons contributes $\pi - \frac{4}{3}\pi = -\frac{\pi}{3}$: whatever the shape of a simply connected flake, the former always exceed the latter by exactly six, so that the boundary term is $2\pi$, just as the pentagons of a fullerene are always twelve.

\begin{table}
\centering
\tbl{Euler characteristic for various examples.\label{tab:VariousEuler}}
{\begin{tabular}{|c|c|c|c|}
  \hline
  SURFACE       & $g$ & $b$ & $\chi$ \\ \hline
  BEL           & 0 & 1 & 1 \\ \hline
  ELL           & 0 & 1 & 1 \\ \hline
  HYP           & 0 & 2 & 0 \\ \hline
  Disk          & 0 & 1 & 1 \\ \hline
  Spherical cap & 0 & 1 & 1 \\ \hline
  Punct. Sphere & 0 & 1 & 1 \\ \hline
  Sphere        & 0 & 0 & 2 \\ \hline
  Cylinder      & 0 & 2 & 0 \\ \hline
  Annulus       & 0 & 2 & 0 \\ \hline
  Torus         & 1 & 0 & 0 \\ \hline
\end{tabular}}
\end{table}

\begin{table}
\centering
\tbl{The five Platonic solids: X-hedra. Here $q$ is the number of edges meeting at each vertex (degree of the vertex).\label{tab:PlatoSolids}}
{\begin{tabular}{|c|c|c|c|c|c|}
  \hline
   & Tetra & Hexa & Octa & Dodeca & Icosa \\ \hline
  $V$ & 4 & 8 & 6 & 20 & 12 \\
  $E$ & 6 & 12 & 12 & 30 & 30 \\
  $F$ & 4 & 6 & 8 & 12 & 20 \\
  $q$ & 3 & 3 & 4 & 3 & 5 \\
  \hline
\end{tabular}}
\end{table}
The formula \eqref{EulPoinPolyhedra} is invariant under the duality transformation
\begin{equation}\label{duality}
  V \to F \,, F \to V \,, E \to E \,,
\end{equation}
for which Hexahedron $\leftrightarrow$ Octahedron and Dodecahedron $\leftrightarrow$ Icosahedron, with the Tetrahedron being self-dual.

The only regular polygons used for the five Platonic solids are: Triangle ($p=3$), Square ($p=4$) and Pentagon ($p=5$). The reason is in the rules that the tessellation/tiling of the surface must obey, as we now explain.

\begin{figure}[htbp]
  \centering
  \subfloat[]{\includegraphics[width=0.19\linewidth]{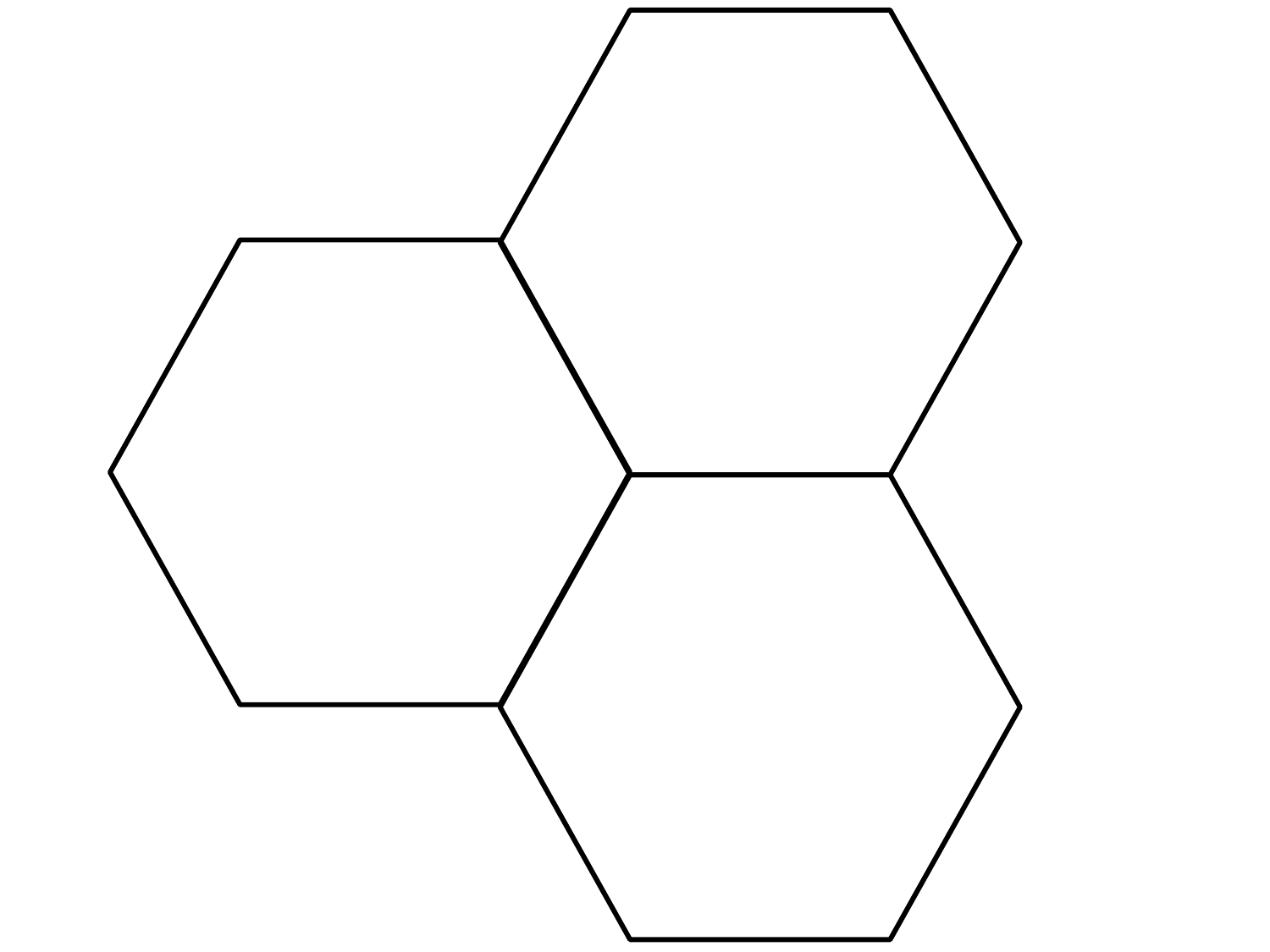}\label{HeTil}}
  \subfloat[]{\includegraphics[width=0.19\linewidth]{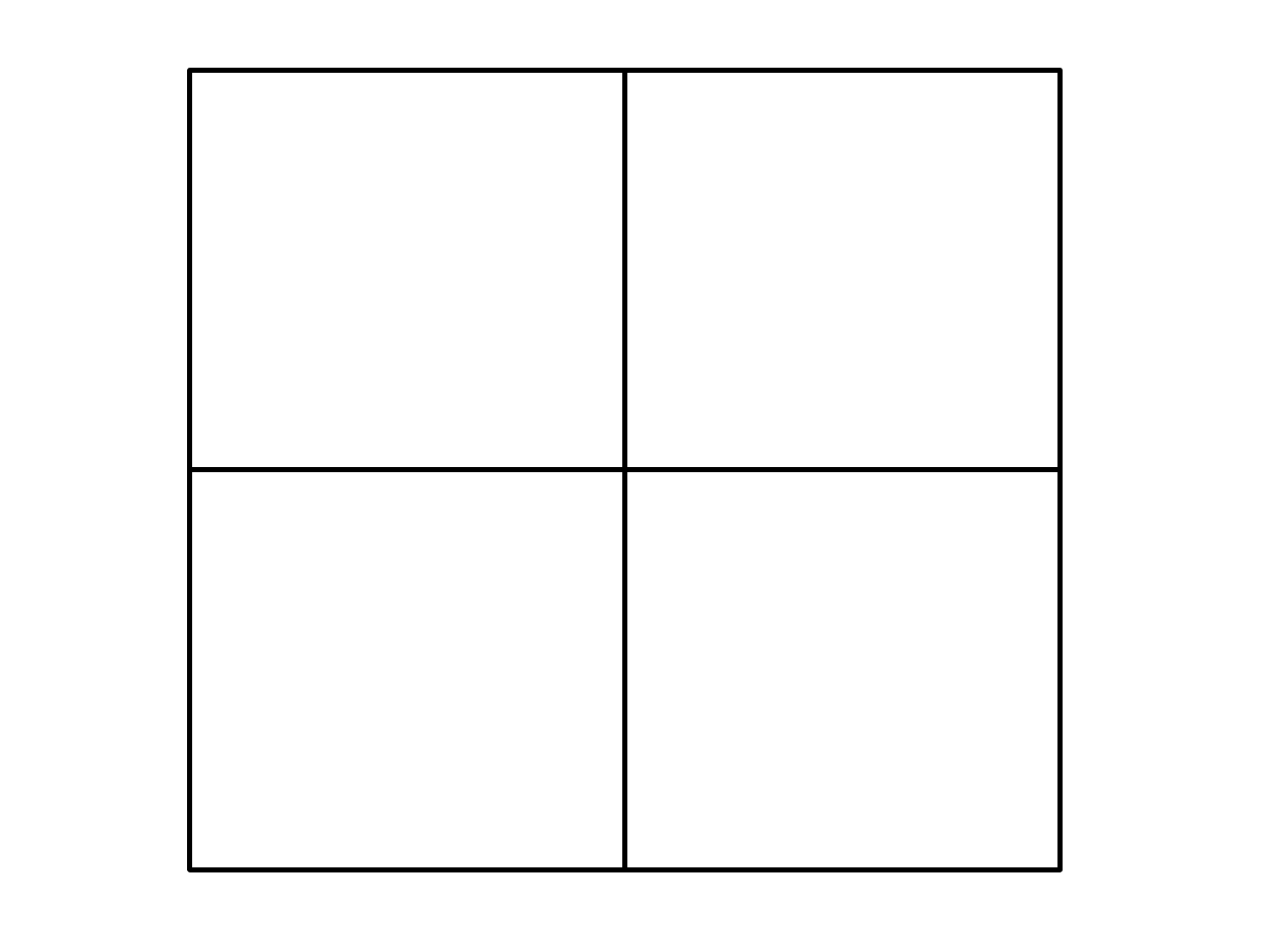}\label{SqTil}}
  \subfloat[]{\includegraphics[width=0.19\linewidth]{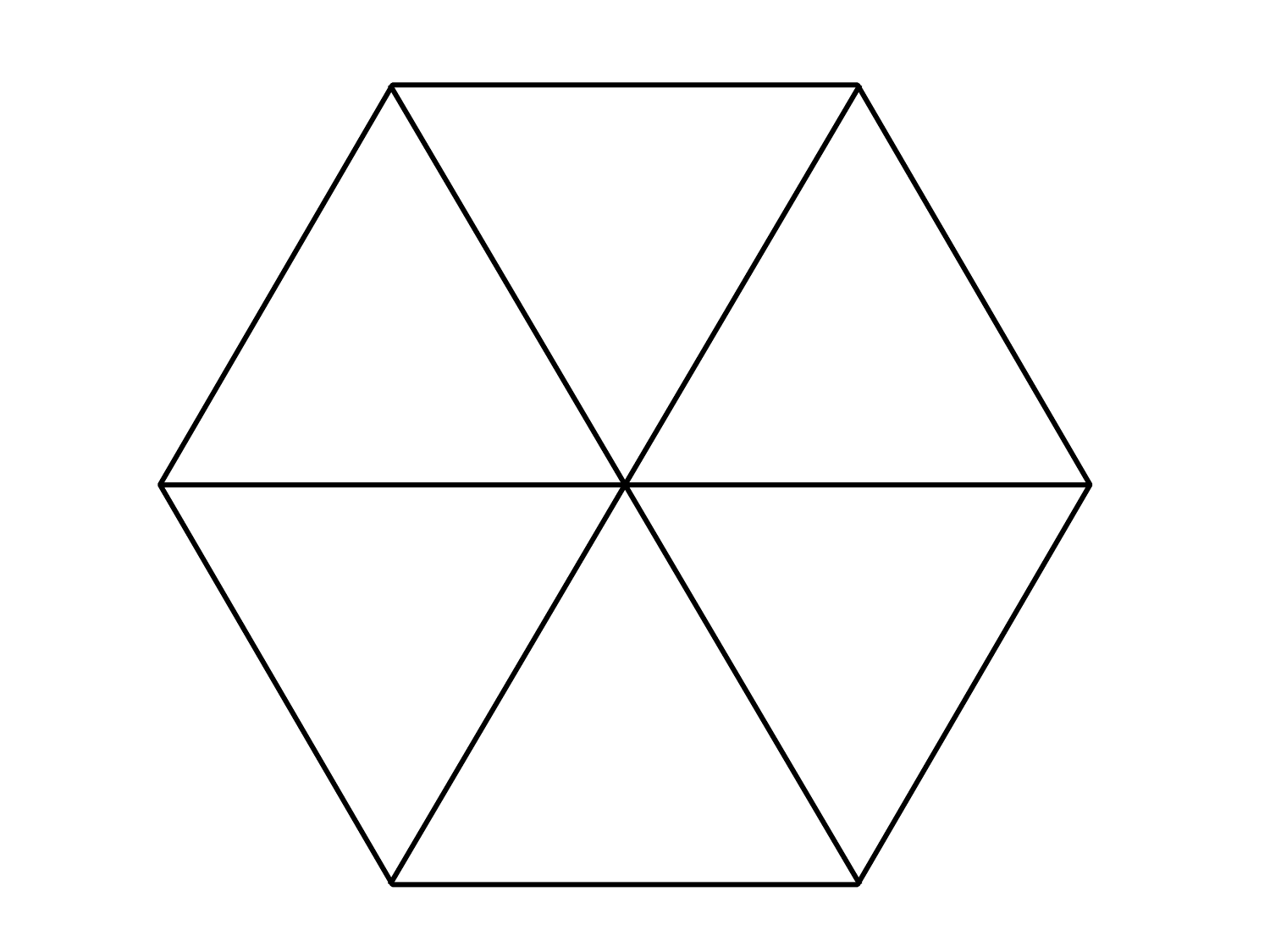}\label{TrTil}}
  \caption{The three cases of Table~\ref{tab:PlaneTessel}.}
  \label{PlaneTessel}
\end{figure}

The simplest case is the tessellation of the plane. Say $q$ the degree of the vertices, $p$ the number of edges of the polygon, $n$ the number of polygons used in the tessellation. They must all be natural numbers. With these $V = (n  p)/q$, $E = (n p)/2$ and $F = n$, hence, using those expressions in the Euler formula for a periodic tessellation of the plane, i.e., of the torus, $V-E+F=0$ (equivalently, for a large patch, $V-E+F=1$, whose contribution per polygon, $1/n$, vanishes as $n\to\infty$), we have
\begin{equation}
 \frac{n \, p}{q} \,  - \frac{ n \, p}{2} + n = 0 \,,
\end{equation}
Notice that the formula gives no constraints on $n$ ($n$ can be dropped), as it must be for the infinite plane, $p/q  - p/2 + 1 = 0$, hence
\begin{equation}\label{pANDqPLANE}
  p = \frac{2 q}{q - 2} \,,
\end{equation}
from which Table~\ref{tab:PlaneTessel} and Fig.~\ref{PlaneTessel} descend

\begin{table}
\centering
\tbl{Combinations of number of sides of the polygon, $p$, and degree of the vertices, $q$, for the three only possible tessellations of the plane with regular polygons.\label{tab:PlaneTessel}}
{\begin{tabular}{|c|c|c|c|}
  \hline
       & $p$ & $q$ &  dual to  \\ \hline
  (a)  & 6   & 3   &  (c)     \\
  (b)  & 4   & 4   & self     \\
  (c)  & 3   & 6   &  (a)     \\
  \hline
\end{tabular}}
\end{table}

Thus, the tiling of the plane with regular polygons can only happen with hexagons, squares, triangles. No other choices of $p,q \in \mathbb{N}$ are possible. Notice also that the hexagon is the only regular polygon made of regular (equilateral) triangles. That is the reason behind the duality, here being $q \leftrightarrow p$, between hexagonal and triangular tessellation of the plane. Clearly the square, $p = 4 = q$, is self-dual.

The aware reader here might wonder how it is possible that a flat surface, the plane, can be tiled with squares and triangles. The reason is in the degree of the vertices of the tessellation. If one insists in having $q=3$ for the squares and the triangles too, then it would not be possible to tile the place with regular squares or triangles, but only with hexagons.

For the sphere
\begin{equation}
 \frac{n \, p}{q} \,  - \frac{ n \, p}{2} + n = 2 \,.
\end{equation}
The formula now depends on $n$, i.e., only for specific numbers of the given regular polygon we can close a sphere. Simple algebra gives
\begin{equation}
  n [(2 - q) p + 2 q] = 4 q \,,
\end{equation}
that gives Table~\ref{SphereTessel}
\begin{table}[htbp]
\centering
\tbl{The only possible tilings of the sphere with regular polygons, from $n[(2-q)p+2q]=4q$: $n$ polygons with $p$ sides, $q$ at each vertex. Duality, $p\leftrightarrow q$, exchanges hexahedron and octahedron, dodecahedron and icosahedron, while the tetrahedron is self-dual.\label{SphereTessel}}
{\begin{tabular}{|c|c|c|c|c|}
  \hline
  $q$ & condition & $p$ & $n$ & solid \\ \hline
  \multirow{3}{*}{3} & \multirow{3}{*}{$n(6-p)=12$} & 3 & 4 & tetrahedron \\
   & & 4 & 6 & hexahedron (cube) \\
   & & 5 & 12 & dodecahedron \\ \hline
  4 & $n(4-p)=8$ & 3 & 8 & octahedron \\ \hline
  5 & $n(10-3p)=20$ & 3 & 20 & icosahedron \\ \hline
\end{tabular}}
\end{table}

Let us now take one more step towards the tessellation of the pseudospheres. Let us consider the tesselation formula, descending from the Euler formula, in this form
\begin{equation}\label{EulerPoincareGeneral}
 \sum_{n} n_p \left( 6 - p \right) = 6 \, \chi ({\cal M}) \,,
\end{equation}
where $n_p$ is the number of regular polygons of $p$ sides and, for simplicity, a definite degree of the vertices is fixed to be $q=3$.

We notice here three facts. \textit{First}, for $p > 6$ the polygons contribute with a negative factor to $\chi$, hence, through the Gauss--Bonnet theorem, to $K$. Henceforth, pairs of ``curvature charges'' of the same absolute value but with opposite sign, e.g., $5-7$ (``curvature charges'': + 1 - 1) or $4-8$ (``curvature charges'': + 2 - 2) or $3-9$ (``curvature charges'': + 3 - 3), are always topologically allowed for any tiling, since they contribute a net zero, overall. \textit{Second}, while 3 is the lower limit for $p$ (if degenerate cases are not considered), there is no upper limit. Henceforth, the negative ``curvature charge'' carried by a polygon, with $p > 6$, can be any.

The reason for that is the noncompact nature of SO(2,1), the symmetry group of the Lobachevsky plane (or of the Poincar\'e disk or of the hyperboloid model), that is the geometry of constant negative Gaussian curvature, $K < 0$ . This group is the negative curvature counterpart of SO(3), the symmetry group of the sphere, $K>0$. Besides the two infinite families of cyclic and dihedral groups, the latter has only three exceptional finite subgroups, the tetrahedral, octahedral and icosahedral groups, that are the symmetry groups of the Platonic solids. The former, being noncompact, has only cyclic finite subgroups (rotations about a point), but it has infinitely many \textit{infinite} discrete subgroups, the Fuchsian groups, and that makes its ``crystallography'' (see Section~\ref{sec:LobachevksyPlatonicSolids}). Preliminary results in this comparison, within the carbon structure laboratory realization, can be found in Ref.~\protect\refcite{LobLab_2016}. \textit{Third}, in the middle is the hexagon, that never carries any ``curvature charge'''. Henceforth, we can add any number of them to the tessellation of any surface.

To schematically summarize our discussion, we recall that there are three geometries: the spherical, $K>0$, the Euclidean, $K=0$, the Lobachevsky, $K<0$. The polygons that best represent each of those geometries, in a tessellation, are: the pentagon, for $K>0$; the hexagon, for $K=0$; the heptagon (and all polygons with $p>6$), for $K<0$. The two missing regular polygons, the equilateral triangle and the square, are the only two that can live in both the $K=0$ and the $K>0$ geometries, but not in the $K<0$ geometry.

\begin{figure}[htbp]
  \centering
  \subfloat[Tetra]{\includegraphics[width=0.19\linewidth]{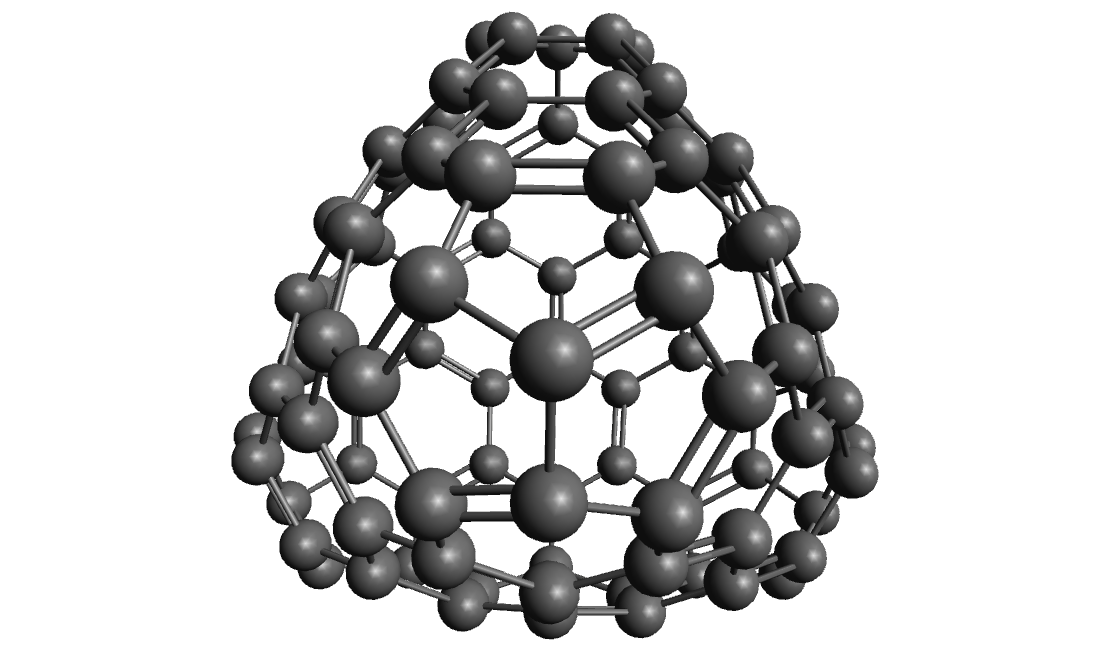}\label{te}}
  \subfloat[Hexa]{\includegraphics[width=0.19\linewidth]{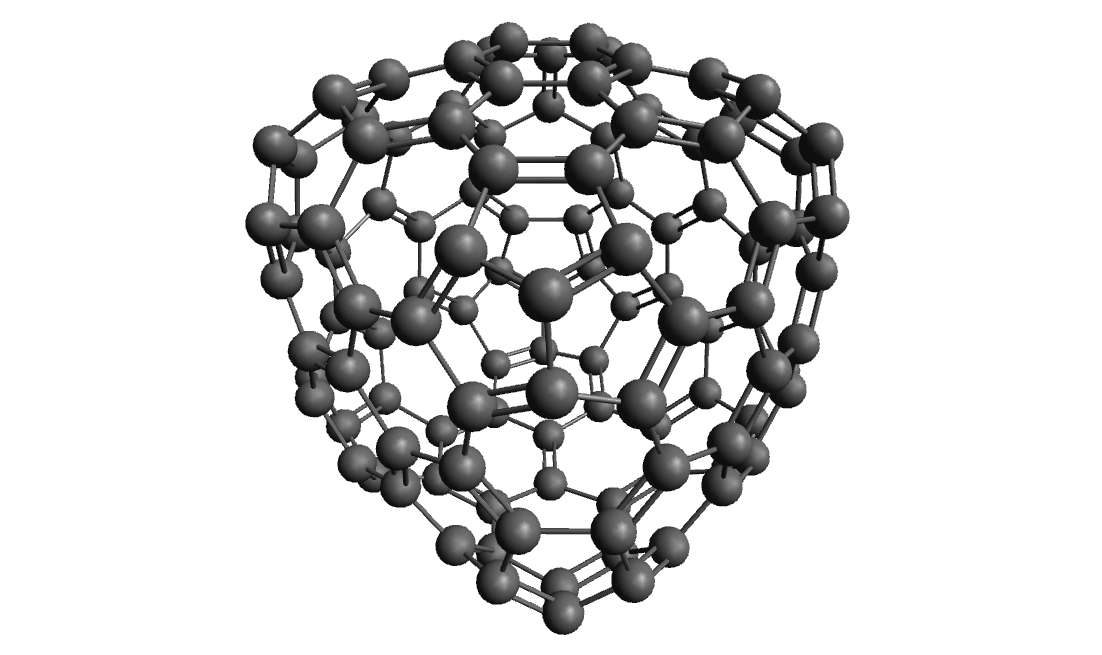}\label{cu}}
  \subfloat[Octa]{\includegraphics[width=0.19\linewidth]{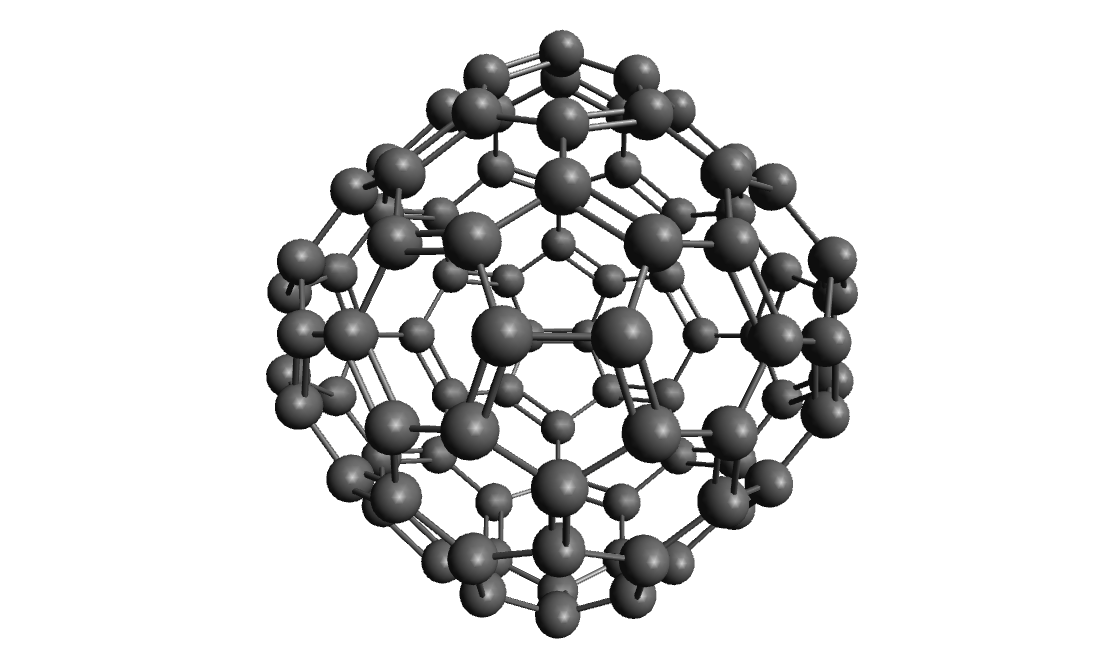}\label{oc}}
  \subfloat[Dodeca]{\includegraphics[width=0.19\linewidth]{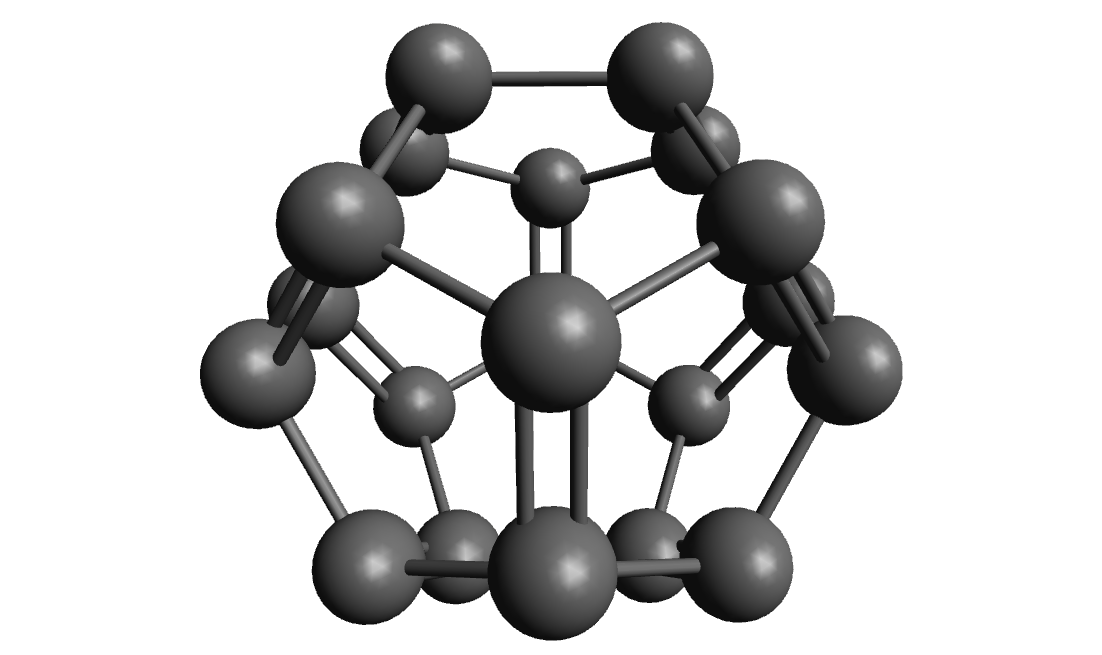}\label{do}}
  \subfloat[Icosa]{\includegraphics[width=0.19\linewidth]{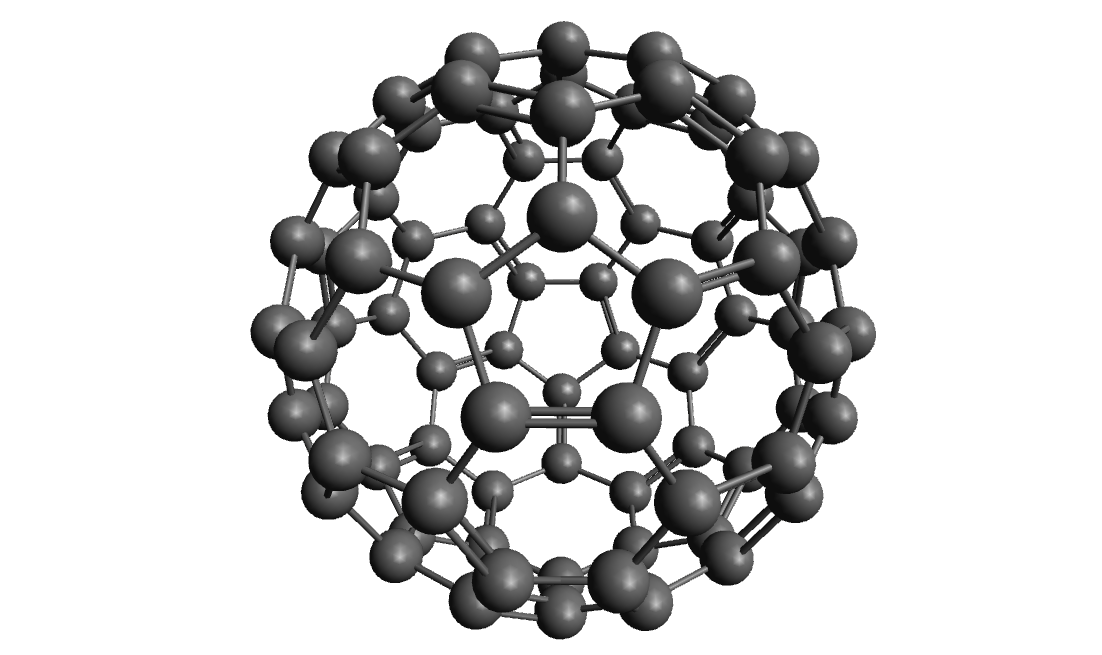}\label{ic}}
  \caption{A look at a possible set of carbon-made X-hedra through one of their vertices: (a) tetrahedron, (b) cube, (c) octahedron, (d) dodecahedron (aka fullerene C20) and (e) icosahedron (aka fullerene C80($I_{h}$)).}
  \label{fig:plasol}
\end{figure}

Fig.~\ref{fig:plasol} shows the Platonic solids made with carbon. Apart from C20, the remaining molecules can only approximate the actual flat-faced polyhedra. Fullerene C80($I_{h}$) is a chamfered dodecahedron. Being the smallest example in an infinite series of icosahedral fullerenes - those equivalent to Goldberg polyhedra - it still appears rather spherical.

\section{Intrinsic curvature without topological defects}\label{sec:DefectFreeCurvature}

A perfect hexagonal lattice has zero intrinsic curvature, $K=0$. To cover with such a lattice a surface of nonzero total intrinsic curvature, when the hexagons are regular and rigid, the only way is to insert disclinations: pentagons (or squares, or triangles) for positive curvature, heptagons (or octagons, etc.) for negative curvature, see, e.g., Refs.~\protect\refcite{SN88,Kleinert,KV92} for the general case, Refs.~\protect\refcite{Bowick:2000aa,bausch2003grain} for the positive curvature case and Refs.~\protect\refcite{LobLab_2016,iorio2014} for the negative curvature case in graphene.

There is, however, a second, and somehow counter-intuitive, way to have intrinsic curvature \cite{Gauss1827,Spivak1979} on a hexagonal lattice: when the hexagons are \textit{elastic}, i.e., their edges can change length and their angles can change amplitude, a hexagonal patch can carry a (vertex by vertex, or even continuous) distribution of Gaussian curvature without any change in connectivity. Such curvature is \textit{non-topological} in two precise senses:
\begin{itemize}
  \item[i.] it is not produced by a change of topology (the Volterra cut-and-glue process), hence it cannot change the topology of the manifold/discrete structure/lattice, and its total contribution to $K_{tot}$, for a given boundary, is fixed by Gauss--Bonnet;
  \item[ii.] it fades away when the structure is relaxed, i.e., it is not a persistent property of the lattice, but rather of the stress state the lattice is in.
\end{itemize}
In contrast, the curvature carried by disclinations survives relaxations and, together with the boundaries, \textit{is} the topology. In a way, this accounts for the name ``topological'' defect.

That Gaussian curvature can be sustained by elastic strain alone is well established in the elasticity theory of two-dimensional crystalline membranes: in the F\"oppl--von K\'arm\'an compatibility equation the Gaussian curvature and the density of disclinations enter as sources of opposite sign, so that defects ``screen'' the curvature, and they are needed only beyond a threshold, see Refs.~\protect\refcite{SN88,Bowick:2000aa,BowickGiomi2009,VitelliNelson2004,VitelliLucksNelson2006,irvine2010pleats,AzadiGrason2014,AzadiGrason2016}. What we add here is the explicit topological bookkeeping of such non-topological curvature, through the Gauss--Bonnet theorem with boundaries, its extremal case (a sphere-like structure made of hexagons only, that is a punctured sphere), and its consequences for the counting of defects on the pseudospheres. We present several independent, although interconnected, arguments: discrete deficit-angle calculations, a continuum Gauss--Bonnet analysis, an explicit polyhedral construction, a nonlinear-metric computation and energetic considerations.

\subsection{Discrete picture: three types of deformations}\label{sec:ThreeDeformations}

We use the deficit angle at a vertex, $\delta_V = 2\pi - \sum_i \theta_i$, see Appendix~\ref{app:DeficitAngles}.

\begin{figure}[htbp]
  \centering
  \includegraphics[width=\linewidth]{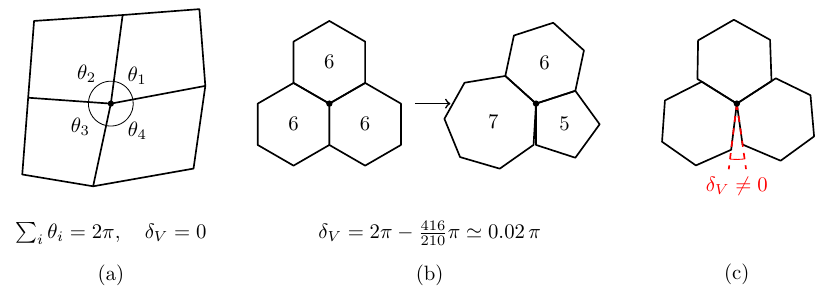}
  \caption{The three types of deformations of Section~\ref{sec:ThreeDeformations}: (a) the individual angles change, but not their sum, hence no curvature; (b) change of connectivity: three regular hexagons replaced by a regular pentagon, hexagon and heptagon, giving a small local deficit angle, while the total curvature of the $5$--$7$ dipole is zero; (c) no change of connectivity: deformed hexagons open a deficit angle, all carried by the hexagons, and the gap closes only when the hexagons buckle out of the plane.}
  \label{fig:deformations}
\end{figure}

\textit{First type.} We can deform the polygons keeping the deficit angle fixed, i.e., we change the individual $\theta_i$s but not their sum, as in Fig.~\ref{fig:deformations}(a) for the plane. These deformations carry no change in local or global curvature, nor in topology.

\textit{Second type.} We can change the connectivity, changing the local curvature, but keeping the total curvature and the topology fixed, as in Fig.~\ref{fig:deformations}(b). The vertex common to three regular hexagons carries $\delta_V = 2\pi - 3\cdot\frac{2}{3}\pi = 0$, whereas the vertex common to a regular pentagon, a regular hexagon and a regular heptagon carries
\begin{equation}\label{DefAngleScar}
  \delta_V = 2\pi - \sum_{n=5}^{7}\frac{n-2}{n}\,\pi = 2\pi - \frac{416}{210}\,\pi \simeq 0.02\,\pi \,.
\end{equation}
Thus regular (rigid) polygons induce a local curvature that must add up to zero when all vertices of the dipole are considered, whereas if the polygons are allowed to deform we can still lie in a plane.

\textit{Third type.} Without changing the connectivity we can still change the local intrinsic curvature, as in Fig.~\ref{fig:deformations}(c): deforming the hexagons opens a deficit angle (or produces an excess angle), and to close the structure the hexagons must buckle out of the plane.

Hence, a smooth pattern of bond-length \textit{and/or} bond-angle variations induces a distributed Gaussian curvature without altering the network connectivity. Notice that keeping the edge lengths fixed does \textit{not} forbid intrinsic curvature: the angles alone can do the job. This is the case of the Chebyshev (fisherman's) nets, where all edges have the same length and only the angles vary, and that do cover curved surfaces, including those of constant negative curvature, whose Chebyshev nets are governed by the sine-Gordon equation \cite{mclachlan}. In the discrete setting, an equilateral but non-equiangular hexagon gives, in general, $\delta_V \neq 0$.

\subsection{Continuum picture: from the disk to the spherical cap}\label{sec:DiskToCap}

The Gauss--Bonnet theorem \eqref{GaussBonnetFull} tells us that one can deform $\cal M$ very drastically, changing its local curvature at every point: as long as the deformations are homeomorphisms (no cuts, no tearing apart, no gluing), the Euler characteristic on the right-hand side is untouched. Notice that, if one needs to count how many defects are necessary for a given surface to have a given total curvature, the first term is the only relevant one. For compact surfaces without boundaries that term carries all the topology; for surfaces with boundaries the topology is shared with the boundary term \cite{Eguchi:1980jx}.

\begin{figure}[htbp]
  \centering
  \includegraphics[width=0.6\linewidth]{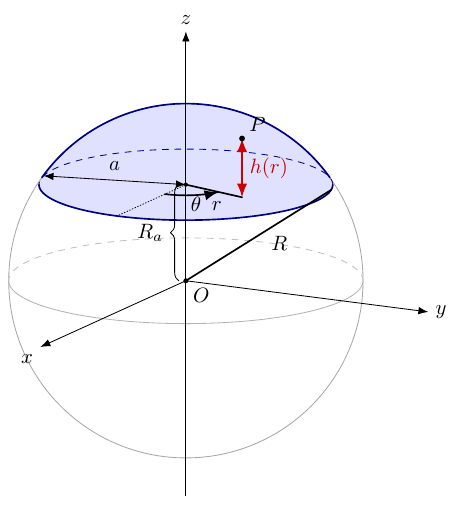}
  \caption{The spherical cap on the sphere of radius $R$ and center $O$, obtained by deforming its base, the planar disk of radius $a$, at distance $R_a = \sqrt{R^2-a^2}$ from $O$. The point at distance $r$ from the axis and azimuth $\theta$ is displaced by $h(r)$ to the point $P$ of the cap: in cylindrical coordinates, $P = (r,\theta,h(r))$. The bulk and boundary terms of the Gauss--Bonnet theorem change, their sum, $2\pi\chi = 2\pi$, does not.}
  \label{fig:diskcap}
\end{figure}

\begin{figure}[htbp]
  \centering
  \subfloat[]{\includegraphics[width=0.8\linewidth]{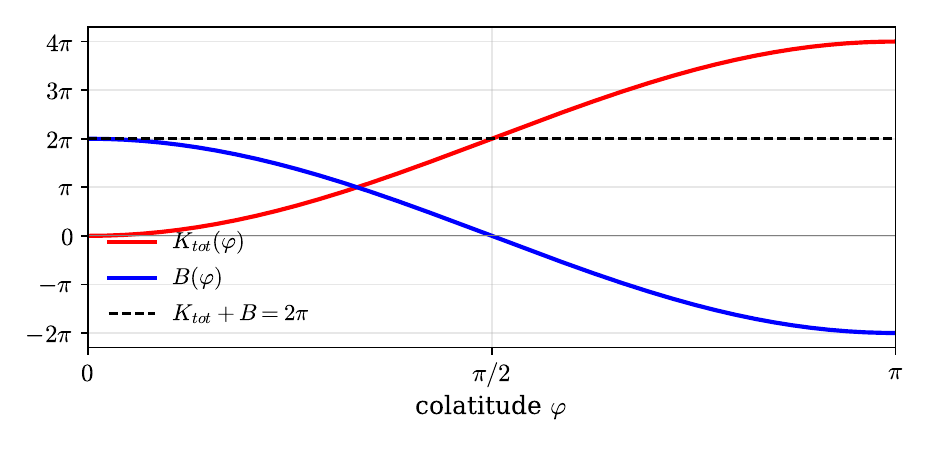}\label{KtotBoundSphericalBound}}\\[1ex]
  \subfloat[]{\includegraphics[width=0.8\linewidth]{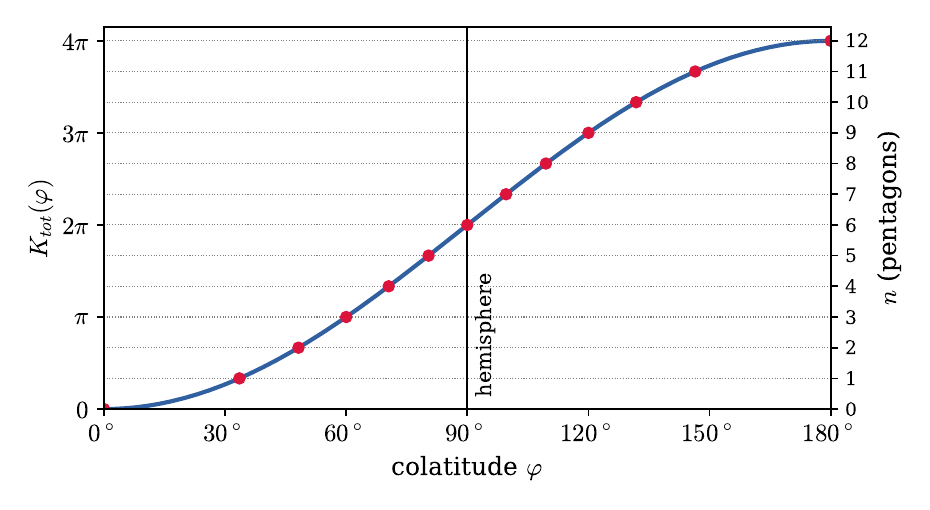}\label{KtotSphericalCup}}
  \caption{(a) Total curvature, $K_{tot}(\varphi)$ (red), and boundary term, $B(\varphi)$ (blue), for the spherical cap of colatitude $\varphi$: their sum is $2\pi$ for all $\varphi$. At the north pole the boundary term makes the whole Euler characteristic, at the equator (a geodesic) the contribution is entirely from the curvature, beyond it the boundary contribution becomes negative and reaches $-2\pi$ at the south pole, where $K_{tot}=4\pi$. (b) $K_{tot}(\varphi)$ with the colatitudes (dots) at which the cap has collected $n\,\pi/3$ of curvature, i.e., the curvature of $n$ pentagons, $n=1,\dots,12$ (right axis).}
\end{figure}

Consider the planar disk of radius $a$, ${\cal M} = \{(r,\theta): 0\le r\le a,\ \theta\in[0,2\pi)\}$, i.e., the base of the cap of Fig.~\ref{fig:diskcap}. It is flat, so $\int_{\cal M} K\,d^2\mu = 0$, while its boundary, a circle of the plane, has $k_g = 1/a$, hence
\begin{equation}\label{ChiDisc}
  2\pi\,\chi = \int_{\partial\cal M}\frac{d\ell}{a} = 2\pi \quad\Rightarrow\quad \chi({\rm disk}) = 1 \,.
\end{equation}
We now deform the disk continuously, through the out-of-plane displacement $h(r)$ of its points, into a spherical cap of radius $R$ and colatitude $\varphi$, with $a = R\sin\varphi$ and $R_a = R\cos\varphi$, Fig.~\ref{fig:diskcap} (regular shapes ease the computation, but any homeomorphic irregular variant gives the same final result). With $K = 1/R^2$ and area $A = 2\pi R^2(1-\cos\varphi)$
\begin{equation}\label{KtotCalotta}
  K_{tot}(\varphi) = \int_{\cal M}K\,d^2\mu = 2\pi(1-\cos\varphi)\,.
\end{equation}
The boundary is a circle of radius $R\sin\varphi$, whose total curvature is $1/(R\sin\varphi)$ and whose normal curvature is $1/R$, hence, from $k^2 = k_g^2 + k_n^2$,
\begin{equation}\label{GeoCurvCircle}
  k_g = \frac{\cos\varphi}{R\sin\varphi} \,,\qquad
  B(\varphi) \equiv \int_{\partial\cal M}k_g\,d\ell = 2\pi\cos\varphi \,.
\end{equation}
Therefore
\begin{equation}\label{GaussBonnetCalotta}
  K_{tot}(\varphi) + B(\varphi) = 2\pi(1-\cos\varphi) + 2\pi\cos\varphi = 2\pi = 2\pi\,\chi(S_\varphi)\,,
\end{equation}
that is $\chi({\rm spherical\ cap}) = 1 = \chi({\rm disk})$, for every $0\le\varphi<\pi$, see Fig.~\ref{KtotBoundSphericalBound}. The bulk curvature grows from $0$ to $4\pi$, while the boundary term decreases from $2\pi$ to $-2\pi$, and their sum is always $2\pi$. At the north pole, $\varphi\to0^+$, the area vanishes and the boundary circle collapses to a point with $\oint k_g\,d\ell\to+2\pi$; at the equator, $\varphi=\pi/2$, the boundary is a geodesic and the whole contribution is from the bulk; at the south pole, $\varphi\to\pi^-$, the area tends to $4\pi R^2$ and the boundary circle collapses with $\oint k_g\,d\ell\to-2\pi$, so that $4\pi-2\pi=2\pi$.

The single missing point is of paramount importance: it carries the contribution $-2\pi$ that keeps the surface a topological disk, not a sphere. To close it, a further $+2\pi$ of curvature must be added in a non-continuous way: that is, by the defects.

It is instructive to read Eq.~\eqref{KtotCalotta} in units of the curvature charge of a pentagon, $\pi/3$: the cap of colatitude $\varphi$ has collected the curvature of $n$ pentagons when $2\pi(1-\cos\varphi_n) = n\pi/3$, i.e., at $\varphi_n \simeq 34^\circ$, $48^\circ$, $60^\circ$, $71^\circ$, $80^\circ$, $90^\circ$, $100^\circ$, $109^\circ$, $120^\circ$, $132^\circ$, $146^\circ$, $180^\circ$, for $n=1,\dots,12$, see Fig.~\ref{KtotSphericalCup}. This is the bridge between the defect-free and the defected descriptions of the same curved surface.

\subsection{A sphere made of hexagons only is a punctured sphere}\label{sec:StereoHexagons}

\begin{figure}[htbp]
  \centering
  \resizebox{0.5\textwidth}{!}{\input{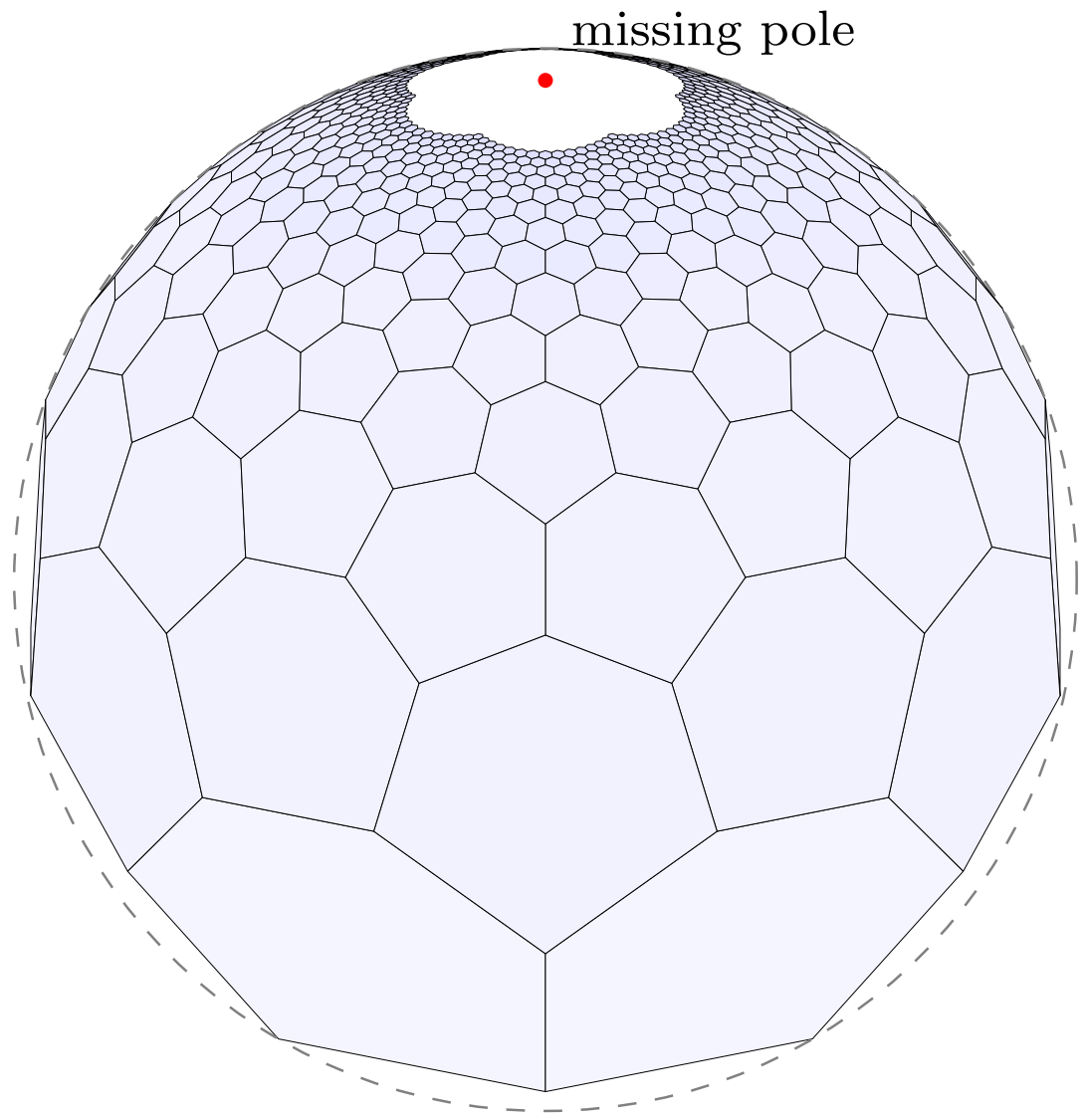}}
  \caption{A sphere-like structure made of hexagons only, whose sizes decrease from the south pole towards the north pole, that is missing: topologically, a punctured sphere, i.e., a disk. It is the inverse stereographic image of the honeycomb lattice (rings $0$ to $14$ shown, the central hexagon is at the south pole, hidden in this view). Each hexagon is planar, all vertices lie on the sphere, and all deficit angles are positive. The hexagons accumulate at the north pole, the projection point, that is missing.}
  \label{fig:HexSphereTikz}
\end{figure}

Topologically, building a sphere-like structure belt after belt of hexagons, starting from one hexagon at a pole, is equivalent to starting from a disk and adding to it, at each step, a strip, obtaining at each step another topological disk, now with nonzero curvature. The belts are necessarily made of \textit{smaller and smaller} hexagons, so that their deficit angles distribute the curvature, Fig.~\ref{fig:HexSphereTikz}.

The combinatorial obstruction to closing the structure is immediate. For any \textit{finite} trivalent cell complex made of hexagons only (however deformed), $3V = 2E$ and $6F = 2E$, so that $\chi = V - E + F = 0$ if closed, see also Eq.~\eqref{EulerPoincareGeneral}, and $\chi=1$ for any simply connected patch: never $2$. Therefore, a hexagons-only ``sphere'' can only be an \textit{infinite} complex, whose hexagons accumulate at a point, that is the missing point.

What is less immediate is the \textit{explicit} computation of the curvatures carried by the various belts of hexagons, and the proof that they add up exactly to $4\pi$, the whole curvature of the sphere, while the missing point accounts for the $-2\pi$ of the boundary. Here is a construction that makes it possible. The vertices of a regular hexagon lie on a circle. The inverse stereographic projection maps circles of the plane into circles of the sphere, and a circle of the sphere lies on a plane of $\mathbb{R}^3$. Hence, the inverse stereographic image of the honeycomb lattice is an infinite polyhedral surface made of \textit{planar} (non-regular, cyclic) hexagons, with trivalent vertices, inscribed in the sphere, whose only missing point is the projection point (the north pole), where the images of the hexagons accumulate, see Fig.~\ref{fig:HexSphereTikz}. This surface is the image of the whole plane and the latter, being homeomorphic to the punctured sphere, has $\chi = 1$. Adding the point at infinity, i.e., the north pole, closes the plane into the sphere, $\chi = 2$: that single point carries exactly the curvature that the rings of hexagons accumulate only in the limit, and it is the discrete counterpart of the $+2\pi$ that, in Section~\ref{sec:DiskToCap}, closes the cap into the sphere. Moreover, as the circumcircle of a hexagon of the honeycomb contains no other vertex of the lattice, and the stereographic projection preserves this property, every face plane leaves all the other vertices on one side: the polyhedral surface is convex and all the deficit angles $\delta_V$ are strictly positive.

\begin{figure}[htbp]
  \centering
  \begin{tikzpicture}
    \begin{semilogxaxis}[width=0.7\textwidth,height=0.42\textwidth,
      xlabel={ring $n$}, ylabel={$\sum_{\rm rings\le n}\delta_V\,/\,\pi$},
      ymin=0, ymax=5, xmin=1, xmax=450, grid=major, legend pos=north west]
      \addplot[thick,blue,mark=*,mark size=1pt] table[x=ring,y=cumpi]{cum_a005_sel.dat};
      \addlegendentry{honeycomb, $\ell/R = 0.05$}
      \addplot[thick,red,dashed,domain=1:450] {4};
      \addlegendentry{$4\pi$}
    \end{semilogxaxis}
  \end{tikzpicture}
  \caption{Cumulative curvature (sum of the deficit angles) of the inverse stereographic image of the honeycomb, ring after ring of hexagons, from the south pole. The sum tends to $4\pi$; the missing $-2\pi$ is carried by the boundary collapsing onto the north pole.}
  \label{fig:CumulativeCurvature}
\end{figure}
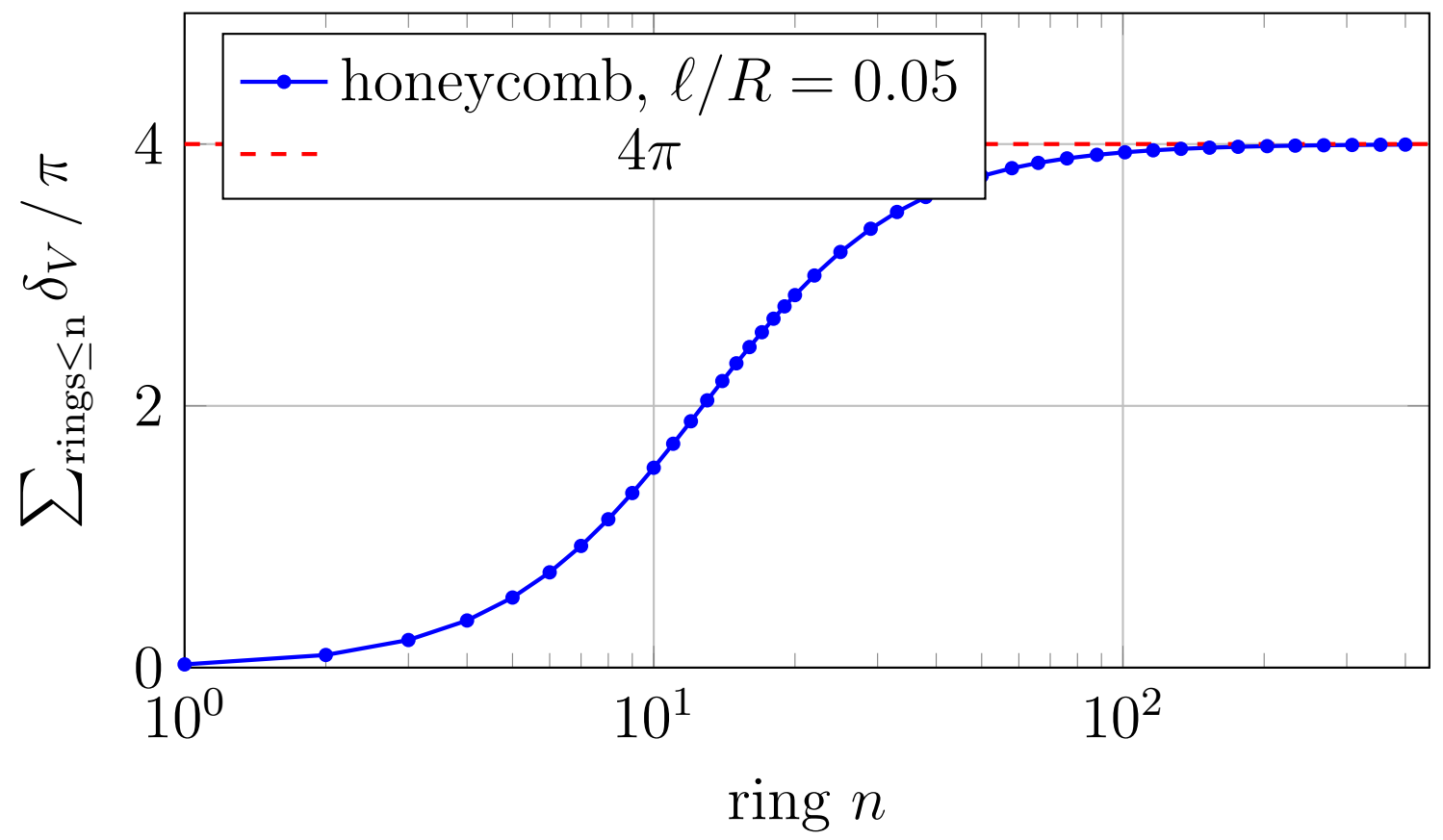

The deficit angles can then be computed vertex by vertex, and summed ring after ring of hexagons, starting from the central hexagon at the south pole. The result of the numerical computation is in Fig.~\ref{fig:CumulativeCurvature} and in Table~\ref{tab:CumulativeCurvature}: the cumulative curvature approaches $4\pi$, the curvature of ring $n$ decays as $n^{-3}$ and the curvature still missing after $n$ rings decays as $n^{-2}$. The limit does not depend on the lattice spacing, while the approach to it does. No finite number of rings reaches $4\pi$: the total is attained only in the limit, i.e., at the missing point. That the limit is exactly $4\pi$ can be shown without summing the series: for any finite patch, the discrete Gauss--Bonnet theorem gives $\sum_{V\in{\rm int}}\delta_V = 2\pi - \tau_{\partial}$, with $\tau_\partial$ the total turning angle of the boundary polygon, and $\tau_\partial \to -2\pi$ as the boundary shrinks around the north pole. This is the discrete counterpart of the limit $\varphi\to\pi^-$ of Eq.~\eqref{GaussBonnetCalotta}. How deformed hexagons produce a deficit angle was made explicit in Ref.~\protect\refcite{vincenzino}, through a ring of hexagons, each with one side $\ell$ and five sides $k\ell$: imposing the closure of each hexagon in the complex plane relates $k$ to the internal angles, and gives a deficit angle at the vertices that grows from $0$ to $\pi/3$ as the hexagons depart from regularity. Extending such a closed-form description to all the rings of the stereographic construction, where the deformation changes from ring to ring, is an open problem of discrete geometry.

\begin{table}[htbp]
\centering
\tbl{Curvature carried by the $n$-th ring of hexagons and cumulative curvature, for the inverse stereographic image of the honeycomb lattice of side $\ell$, on the sphere of radius $R$, $\ell/R=0.05$. Computed with the script \texttt{hexsphere.py} (available with this work).\label{tab:CumulativeCurvature}}
{\begin{tabular}{|c|c|c|}
    \hline
    ring $n$ & curvature of the ring & cumulative curvature$/\pi$ \\ \hline
    1   & $7.7\times10^{-2}$ & 0.025 \\
    10  & $0.61$             & 1.527 \\
    50  & $3.0\times10^{-2}$ & 3.756 \\
    100 & $4.0\times10^{-3}$ & 3.936 \\
    200 & $5.1\times10^{-4}$ & 3.984 \\
    400 & $6.4\times10^{-5}$ & 3.996 \\ \hline
  \end{tabular}}
\end{table}

We conclude that we cannot make a sphere with hexagons only, but only a punctured sphere, and extreme conditions are needed for the hexagons, as they must become of infinitesimal size. Even in this extreme case, with one single point missing, we have not managed to change the topology through the sole deformation of the hexagons, but only the geometry.

\subsection{Nonlinear metric and distributed curvature}\label{sec:NonlinearMetric}

Let us now look at the same issue from the continuum, metric point of view. An out-of-plane displacement $h(x,y)$ of the material point $(x,y)$ induces the metric
\begin{equation}\label{inducedmetric}
  g_{ij} = \delta_{ij} + \partial_i h\,\partial_j h \qquad (i,j=1,2)\,.
\end{equation}
In the linearized limit, $\|\partial h\|\ll1$, one recovers the flat metric ($K=0$), while retaining the nonlinear term $\partial_i h\,\partial_j h$ produces a genuine intrinsic distortion. Notice that, by Gauss' \textit{theorema egregium}, such a displacement, at fixed $(x,y)$, necessarily stretches the material: lengths measured along the membrane change. This is precisely why it can produce intrinsic curvature, and why such curvature costs stretching energy (see Section~\ref{sec:Energetics}).

For instance, for the spherical cap of Fig.~\ref{fig:diskcap},
\begin{equation}\label{hrcap}
  h(r) = \sqrt{R^2-r^2}-\sqrt{R^2-a^2}\,,\qquad r\in[0,a]\,,
\end{equation}
and the first and second fundamental forms of the embedding $(r\cos\theta,\, r\sin\theta,\, h(r))$ give
\begin{equation}\label{Gaussh}
  K = \frac{h'(r)\,h''(r)}{r\,[1+h'(r)^2]^2} = \frac{1}{R^2} \,,
\end{equation}
as it must be: the nonlinear term $\partial_i h\,\partial_j h$, obtained through elastic deformations only, does produce a nonzero \textit{intrinsic} curvature, while its linearization would give $K=0$.

\subsection{Energetics: when do defects take over?}\label{sec:Energetics}

How much curvature can actually be stored in a hexagonal membrane without defects? In the small-slope F\"oppl--von K\'arm\'an description of a two-dimensional crystal of Young modulus $Y$, the Airy stress function $\chi_A$ obeys \cite{SN88,BowickGiomi2009}
\begin{equation}\label{FvK}
  \frac{1}{Y}\nabla^4\chi_A = s(\mathbf x) - K(\mathbf x)\,,
\end{equation}
where $s(\mathbf x) = \sum_\alpha s_\alpha\,\delta(\mathbf x - \mathbf x_\alpha)$ is the density of disclinations of Frank angles $s_\alpha$ ($s=\pm\pi/3$ for a pentagon/heptagon) and $K$ is the Gaussian curvature. The two sources enter with opposite signs: curvature can be ``screened'' by defects, but it can also be sustained by stress alone. For a disk of radius $W$ with uniform $K$ and a disclination $s$ at its center, with stress-free boundary, solving Eq.~\eqref{FvK} gives the elastic energy
\begin{equation}\label{EnergyCapDiscl}
  E(s) = \frac{Y W^2}{32\pi}\,s^2 \;-\; \frac{Y K W^4}{64}\,s \;+\; \frac{\pi\,Y K^2 W^6}{384}\,.
\end{equation}
The last term is the cost of the defect-free curved state, $E(0)\propto Y K^2W^6$; the first term is the energy of a disclination in a flat disk, that grows as $W^2$ \cite{SN88} (not logarithmically, as for a dislocation); the middle term is the defect-curvature interaction. A disclination of charge $s$ lowers the energy, $E(s)<E(0)$, when
\begin{equation}\label{threshold}
  \int K\,d^2\mu \simeq \pi K W^2 > 2 s \,,
\end{equation}
i.e., for a pentagon, when the integrated curvature exceeds $2\pi/3$, that is one sixth of the sphere. This agrees with the threshold found in Ref.~\protect\refcite{AzadiGrason2016}, where it is also shown that dislocation scars lower it to about one twelfth. The threshold is essentially \textit{geometric}: it is set by the integrated curvature, i.e., by the ratio $W/R$ (or by the aspect ratio of a bump \cite{VitelliNelson2004}), and only weakly by the lattice spacing, through the core energies. Below threshold, curvature is accommodated by elastic stress without defects; above it, disclinations (or scars) become energetically favorable, restoring the defect-based description.

When the membrane can also bend, the bending energy (Helfrich) $E_{bend}\simeq \frac{1}{2}\kappa\int(2H)^2 d^2\mu\sim\kappa\,W^2/R^2$ \cite{Helfrich1973,LandauElast} competes with the stretching energy $E(0)\sim Y W^6/R^4$, and the two are comparable at $W^*\sim(\kappa/Y)^{1/4}R^{1/2}$, the geometric mean of the elastic length $\sqrt{\kappa/Y}$ and the radius of curvature. For graphene the bending rigidity $\kappa$ is not negligible: it stems from the misalignment of the $\pi$ orbitals of neighboring atoms, and it sets $\sqrt{\kappa/Y}$ at the sub-\AA ngstr\"om scale.

\subsection{Experimental observations and consequences for the pseudospheres}\label{sec:DefectFreeExperiments}

Graphene nanobubbles grown on the Pt(111) surface are highly strained, curved, defect-free regions, where the strain produces pseudomagnetic fields above $300$\,T \cite{Levy2010}; more generally, wrinkles, ripples and crumples of graphene carry curvature and strain without disclinations \cite{Deng:2016aa}, and the metric induced by strain enters the Dirac equation of the quasiparticles \cite{Vozmediano2010}. On the other hand, in closed shells, or above the threshold \eqref{threshold}, defects do nucleate, as seen in spherical colloidal crystals \cite{bausch2003grain} and in crystals on negatively curved surfaces \cite{irvine2010pleats}.

For the purposes of this work, the important lessons are two. First, graphene shaped as a curved surface between defects does not need to be flat there: it can keep some curvature, through deformed hexagons, e.g., when it is fixed on a substrate. Hence scars, that could be energetically costly, are not the only way to distribute curvature, which can be evenly spread even without defects. Second, such curvature cannot change the topology of the surface, i.e., it cannot change the total Gaussian curvature for given boundaries ($4\pi$ for $S^2$, $-2\pi$ for BEL, etc.). If the defects necessary for the given topology are already there, the extra \textit{local} curvature stemming from the deformed hexagons is bound to contribute zero to the total (\textit{global}) curvature. A general proof of this fact, for arbitrary deformations of the hexagons in the presence of disclinations, is an open problem; for the scars it is a simple counting of 5s and 7s, that need be equal in number.

\section{Lobachevsky surfaces} \label{sec:LobachevksyPlatonicSolids}

We know that there are only five Platonic solids, and that their symmetry groups are the finite subgroups of $SO(3)$, the isometry group of the sphere. What is their counterpart when the curvature is negative? This is the question that motivated \cite{LobLab_2016}, and we now set up the stage to answer it.

\subsection{The Lobachevsky plane and its discrete symmetries}

The Lobachevsky plane can be represented as the upper half-plane $\{(\tilde x,\tilde y)\,|\,\tilde y>0\}$, with the line element
\begin{equation}\label{LobPlane}
  dl^2 = \frac{r^2}{\tilde y^2}\left(d\tilde x^2 + d\tilde y^2\right) \,,
\end{equation}
of constant Gaussian curvature $K = -1/r^2$. Its geodesics are semicircles orthogonal to the ``absolute'', the $\tilde x$-axis, and vertical half-lines. Its (orientation-preserving) isometries are the M\"obius transformations $w\mapsto (aw+b)/(cw+d)$, $w = \tilde x + i\tilde y$, with real coefficients and $ad-bc=1$, i.e., the group $PSL(2,\mathbb{R})$, locally isomorphic to $SO(2,1)$, the Lorentz group in $2+1$ dimensions \cite{LobLab_2016}. This is the negative-curvature counterpart of $SO(3)$.

The crucial difference is that $SO(3)$ is compact, while $SO(2,1)$ is not. The finite subgroups of $SO(3)$ are few (cyclic, dihedral, tetrahedral, octahedral and icosahedral) and they are the point groups of the Platonic solids (only some of them are crystallographic: the icosahedral group is not, as quasicrystals remind us). The finite subgroups of $PSL(2,\mathbb{R})$ are instead only cyclic, while its discrete subgroups, the Fuchsian groups \cite{Katok1992,Magnus1974} (or, including orientation-reversing elements, the non-Euclidean crystallographic groups), are infinitely many, and infinite. The cleanest way to see the contrast is through the triangle groups $\Delta(l,m,n)$ \cite{Magnus1974,Beardon1983}, generated by the rotations about the vertices of a geodesic triangle with angles $\pi/l$, $\pi/m$, $\pi/n$: they are finite, and spherical, when $1/l+1/m+1/n>1$, i.e., for $(2,2,n)$ (dihedral) and $(2,3,3)$, $(2,3,4)$, $(2,3,5)$ (the rotation groups of the Platonic solids); Euclidean, when the sum is $1$, i.e., for $(2,3,6)$, $(2,4,4)$, $(3,3,3)$ (wallpaper groups); and hyperbolic, infinite and Fuchsian, for the infinitely many triples with $1/l+1/m+1/n<1$. For instance, the regular tilings $\{p,q\}$ ($p$-gons, $q$ at each vertex) are spherical, Euclidean or hyperbolic when $1/p + 1/q$ is larger than, equal to, or smaller than $1/2$, respectively. For trivalent tilings, $q=3$, as for graphene, the spherical cases are $p=3,4,5$ (tetrahedron, cube, dodecahedron), the Euclidean one is $p=6$ (the honeycomb), and \textit{all} $p\ge7$ are hyperbolic. The heptagonal tiling $\{7,3\}$ is the closest hyperbolic relative of the honeycomb, and its symmetry group, the $(2,3,7)$ triangle group, plays for $K<0$ the role that the icosahedral group $(2,3,5)$ plays for the dodecahedron and for fullerenes. While the spherical $(2,3,p)$ groups stop at $p=5$, the five Platonic solids, the hyperbolic ones go on for all $p\ge7$: there are infinitely many ``Lobachevsky Platonic solids'', the regular tilings $\{p,q\}$ with $(p-2)(q-2)>4$, each with its own infinite Fuchsian symmetry group. These groups are the basis of the recent crystallography and band theory of hyperbolic lattices \cite{Maciejko2021HyperbolicBand,Boettcher2022Crystallography}, realized in circuit quantum electrodynamics \cite{Kollar2019Hyperbolic}, and were used earlier to describe liquid-crystalline and minimal-surface structures through the crystallography of the Lobachevsky plane \cite{SadocCharvolin1990,SadocMosseri1999}.

\subsection{Hilbert theorem and the infinitely many Lobachevsky surfaces}

There is, however, a fundamental obstruction to realize this geometry in our Euclidean space: by Hilbert's theorem, no complete smooth surface of constant negative curvature can be embedded in $\mathbb{R}^3$ \cite{ovchinnikov}. Every surface of constant $K<0$ in $\mathbb{R}^3$ is only locally isometric to the Lobachevsky plane, it realizes only a portion of it (a ``stripe'', or horocyclic sector), and it carries some unavoidable singularity: cusps, conical points, self-intersections, singular boundaries. Intuitively, these singularities stem from forcing an intrinsically non-compact structure, with symmetry $SO(2,1)$, into a space whose isometries form the compact group $SO(3)$ \cite{LobLab_2016}. The portion realized is, however, larger than it looks. On BEL and HYP there are geodesics that wind many times around the axis: lifted to the Lobachevsky plane, each turn crosses a new copy of the fundamental strip of the surface, so that a geodesic with many windings explores a correspondingly larger piece of the Lobachevsky plane, one strip per turn. This is well known for BEL, whose isometric map onto a horocyclic strip of the upper half-plane makes it explicit \cite{eisenhart1909treatise,iorio2015}, and the near-critical trajectories that wind many times around the throat of the carbon BTZ wormhole \cite{iorio2026classical} are of this kind. As there are infinitely many ways to cut a stripe out of the Lobachevsky plane, there are infinitely many such surfaces, the ``Lobachevsky surfaces'' \cite{ovchinnikov,mclachlan}, related to the solutions of the sine-Gordon equation \cite{mclachlan}: the pseudospheres, the Dini surfaces, the Kuen surface, the breathers, etc. The conjecture of Ref.~\protect\refcite{LobLab_2016} is that this infinity of surfaces is in correspondence with the infinity of discrete subgroups of $SO(2,1)$, suitably adapted to the embedding in $\mathbb{R}^3$.

For tessellations, all this means that there are infinitely many polygons that can carry negative curvature (any $p>6$), that different Lobachevsky surfaces have different Euler characteristics and boundaries, and, through the Gauss--Bonnet theorem, need different numbers and arrangements of defects. The boundary term, which for the sphere is absent, becomes here unavoidable, and it is one of the three actors of the balanced Gauss--Bonnet theorem.

\subsection{Elliptic, parabolic and hyperbolic: the three pseudospheres}\label{sec:EllParHyp}

The elements of $PSL(2,\mathbb{R})$ come in three classes, according to the absolute value of their trace \cite{Katok1992,Beardon1983}: \textit{elliptic} ($|{\rm tr}|<2$), with one fixed point inside the Lobachevsky plane (rotations about that point); \textit{parabolic} ($|{\rm tr}|=2$), with one fixed point on the absolute (translations along horocycles, e.g., $\tilde x\mapsto\tilde x + b$); \textit{hyperbolic} ($|{\rm tr}|>2$), with two fixed points on the absolute (translations along the geodesic joining them, e.g., $\tilde y\mapsto\lambda\tilde y$, $\tilde x\mapsto\lambda\tilde x$). Correspondingly, there are three kinds of one-parameter subgroups.

A surface of revolution of constant $K<0$ has a one-parameter group of isometries, the rotations $v\mapsto v+\phi$ about its axis. Lifted to the Lobachevsky plane, these rotations form a one-parameter subgroup of $PSL(2,\mathbb{R})$, and the three classes give precisely the three pseudospheres of the next section:
\begin{itemize}
  \item \textbf{ELL}: the line element $du^2 + c^2\sinh^2(u/r)\,dv^2$ is that of geodesic polar coordinates about a point, with polar angle $(c/r)\,v$. The rotation about the axis is \textit{elliptic}, and the fixed point is the tip of ELL, a conical point of total angle $2\pi c/r<2\pi$.
  \item \textbf{BEL}: the line element $du^2 + c^2 e^{2u/r}\,dv^2$ is that of horocyclic coordinates. The rotation is \textit{parabolic}, and the fixed point, on the absolute, is the tip of BEL, at infinite distance (a cusp).
  \item \textbf{HYP}: the line element $du^2 + c^2\cosh^2(u/r)\,dv^2$ is that of Fermi coordinates about a geodesic, with arclength $c\,v$ along it. The rotation is \textit{hyperbolic}, and the invariant geodesic is the waist of HYP, a closed geodesic of length $2\pi c$.
\end{itemize}
Thus the names elliptic, parabolic and hyperbolic of the classic differential geometry of pseudospheres are, after all, the names of the conjugacy classes of $SL(2,\mathbb{R})$ to which their rotations belong, and the global invariants that distinguish the three surfaces are the cone angle (ELL), the cusp (BEL) and the length of the closed geodesic (HYP).

This gives the common structure of the discrete symmetries of the defects of the carbon pseudospheres: the rotations $v\mapsto v + 2\pi/p$, that form a cyclic subgroup $\mathbb{Z}_p$ of the elliptic, parabolic or hyperbolic one-parameter group, combined with discrete steps along the meridian, that place the defects at the radii where the surface has accumulated an integer number of units of curvature. We shall see this at work for BEL (Section~\ref{sec:BEL}, \cite{LobLab_2016}), for the family of ELLs (Section~\ref{sec:ELL}), and as a proposal for HYP (Section~\ref{sec:HYP}).

\section{Pseudospheres}\label{PseudospheresMathsGeneral}

In Section~\ref{sec:EllParHyp} the three pseudospheres appeared through the symmetry they keep: they are the Lobachevsky surfaces that are also surfaces of revolution, and their rotation about the axis is an elliptic (ELL), parabolic (BEL) or hyperbolic (HYP) one-parameter subgroup of $SL(2,\mathbb{R})$. Among the infinitely many surfaces of constant $K<0$ \cite{ovchinnikov,mclachlan} (see also Ref.~\protect\refcite{eisenhart1909treatise} and, e.g., the discussion in Ref.~\protect\refcite{iorio2015}), they are the most precise counterparts of the sphere that the Hilbert theorem allows: whereas for the sphere of radius $R$ the Gaussian curvature $K=+1/R^2$ is constant and positive, for a pseudosphere $K=-1/r^2$ is constant and \textit{negative}. We now give their explicit form, which is what is needed to place the defects and to compute the Gauss--Bonnet balance, and which is also the starting point of the applications of Section~\ref{sec:Applications}. From here on, $r$ denotes the (fixed) radius of curvature of the Lobachevsky geometry, as in Eq.~\eqref{LobPlane}, while $R$ denotes either the radius of the sphere (Sections~\ref{sec:TopologicalSurfaces} and~\ref{sec:DefectFreeCurvature}) or, from now on, the radius $R(u)$ of the parallels of a surface of revolution.

Surfaces of revolution in $\mathbb{R}^3$ are typically defined by making a complete turn of $2 \pi$ around, say, the $z$-axis, of a curve $\gamma:[a,b]\rightarrow\mathbb{R}^2$ in the, say,  $(x,z)$-plane. Such surfaces can be parameterized by the inner coordinates $(u,v)$, where $v\in [0,2\pi]$ is the angular coordinate and $u\in[a,b]$ is the meridian coordinate along the surface (see e.g. Ref.~\protect\refcite{eisenhart1909treatise}). \par
	
The embedding of the surface of revolution into $\mathbb{R}^3$ is then given as

\begin{equation}\label{emb}
x(u,v)=R(u)\cos v, \ \ y(u,v)=R(u)\sin v, \ \ z(u)=\pm\int^{u}\sqrt{1-{R'}^2} d\bar{u},
\end{equation}

where $R'= dR/d\bar{u}$, and the function $R(u)$, with its range and domain, fully identifies the surface. The line element is $dl^2=du^2+R^2(u)\,dv^2$, and the Gaussian curvature is $K=-R''(u)/R(u)$ \cite{eisenhart1909treatise}. For $K = - 1/r^2$ the general solution is $R (u) = c_1 \cosh (u/r) + c_2 \sinh (u/r)$, and with an appropriate choice of the constants one finds the three pseudospheres \cite{eisenhart1909treatise}
\begin{align}
\text{ELL:} \ \ \ \ &R(u)=c \ \sinh(u/r)\,, & &R\in[0,r\cos\theta]\,, \label{ELL} \\
\text{BEL:} \ \ \ \ &R(u)=c \ e^{u/r}\,,    & &R\in[0,r]\,, \label{BEL} \\
\text{HYP:} \ \ \ \ &R(u)=c \ \cosh(u/r)\,, & &R\in[c,\sqrt{r^2+c^2}]\,, \label{HYP}
\end{align}
where, for ELL only, $c=r\sin\theta$, and the domains are $u\in[0, r \, {\rm arcsinh}\cot\theta]$ (ELL), $u\in[-\infty, r \ln (r/c)]$ (BEL), $|u|\le r \, {\rm arccosh} \sqrt{1+r^2/c^2}$ (HYP). Here $c$ is a constant that plays a different geometrical role in each pseudosphere (in the correspondence with the BTZ black hole of Section~\ref{sec:BTZ_spectrum} it is related to the horizon, hence to the mass). The minimum of $R$ is the tip of ELL ($R=0$ at $u=0$), the cusp of BEL ($R=0$ at $u=-\infty$) and the waist of HYP ($R=c$ at $u=0$). The maximum is reached at the boundaries, that are necessarily present due to the Hilbert theorem \cite{ovchinnikov}, and that are named ``Hilbert horizons'' in Ref.~\protect\refcite{iorio2014}: ELL and BEL have one each, HYP, whose $R(u)$ is even, has two. The three surfaces are shown in Fig.~\ref{fig:ELLBELHYP}.

\subsection{Limiting cases and Gauss--Bonnet balance}\label{sec:PseudoLimits}

\begin{figure}
    \centering
    \includegraphics[width=0.20\linewidth]{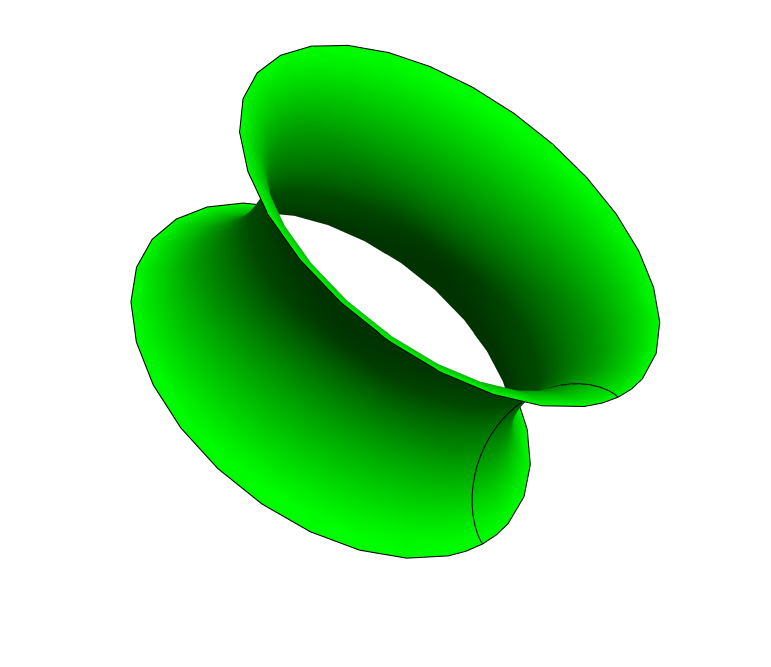}
    \includegraphics[width=0.25\linewidth]{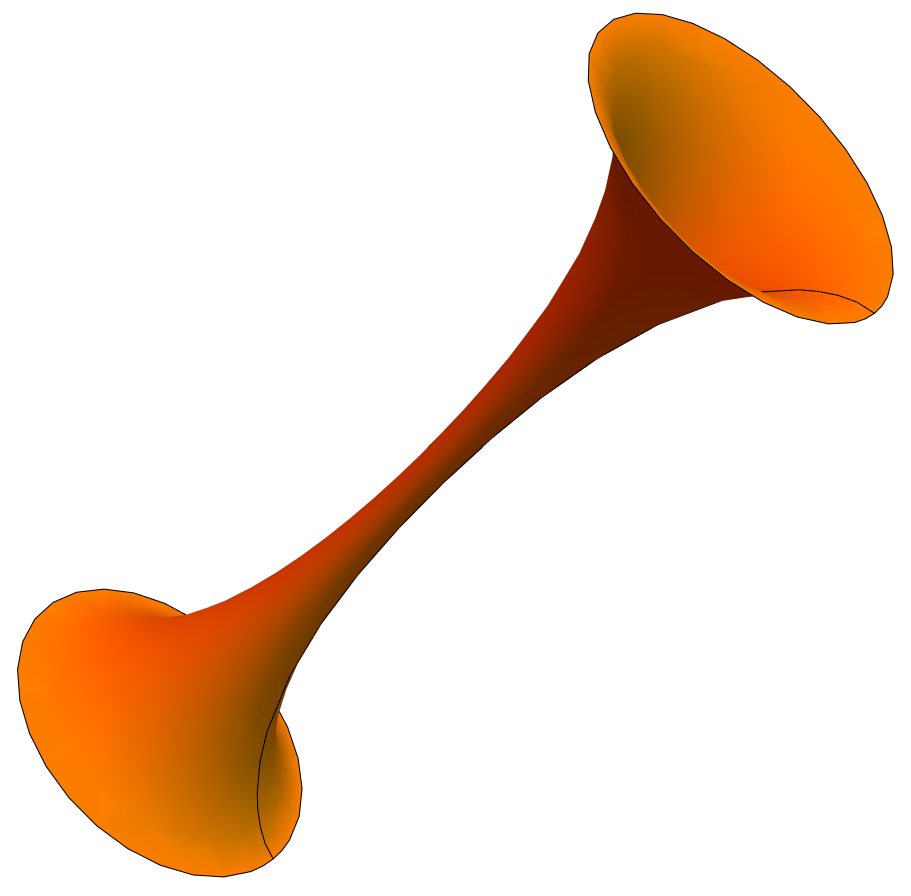}
    \includegraphics[width=0.45\linewidth]{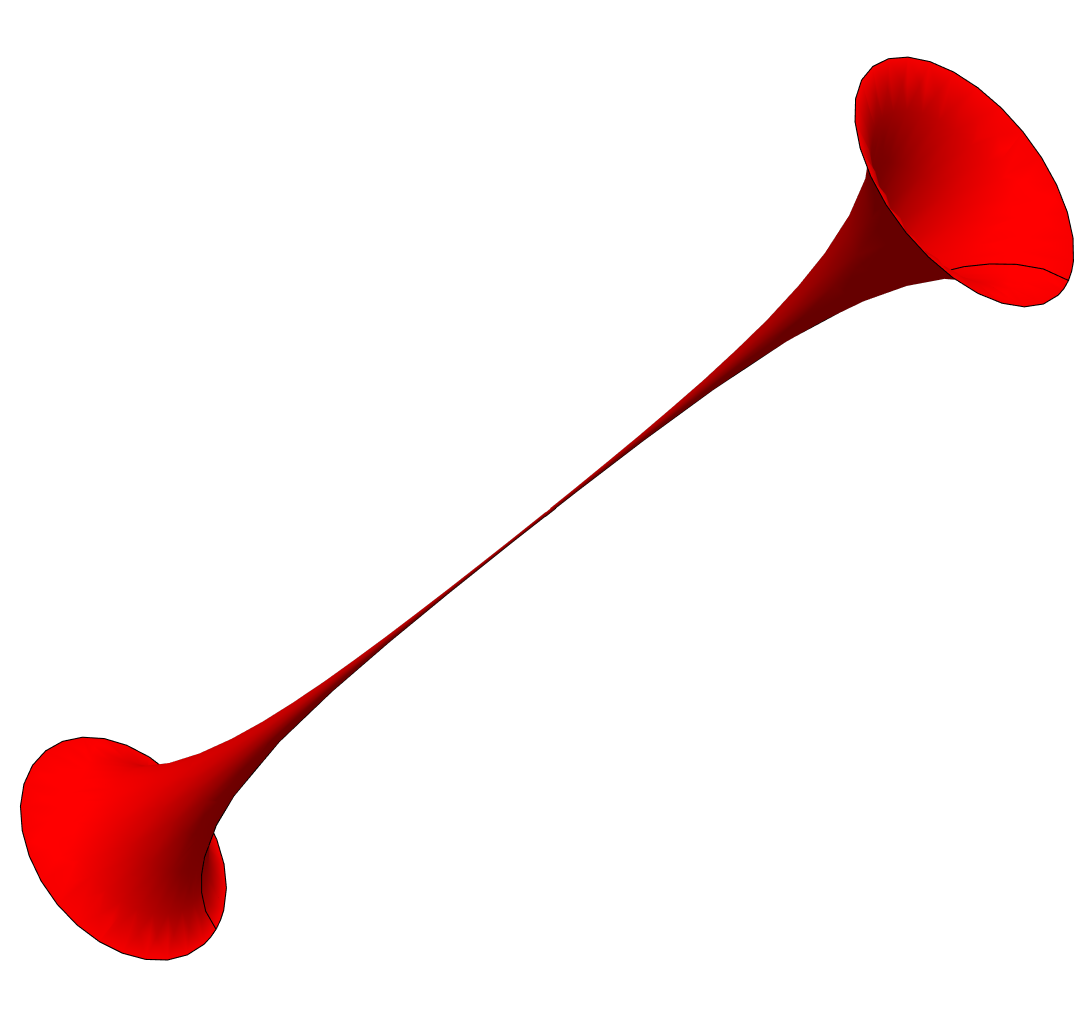}
\caption{HYPs with parameters (left) $r$ = 1 and $c$ = 1, (center) $r$ = 1 and $c$ = 1/10, (right) $r$ = 1 and $c$ = 1/100: for $c/r\to0$ the two halves approach two BELs away from the waist, but a neck of radius $\simeq c$ and length $\simeq 2.6\,r$ survives around it (see Section~\ref{sec:PseudoLimits}).
}
    \label{fig:HYP}
\end{figure}

\begin{figure}
    \centering
    \includegraphics[width=0.30\linewidth]{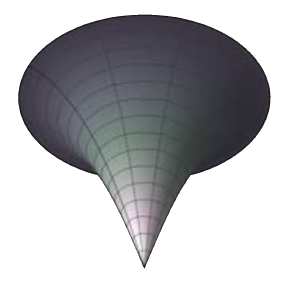}
    \includegraphics[width=0.30\linewidth]{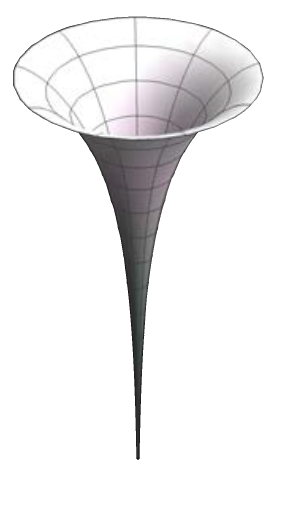}
    \includegraphics[width=0.30\linewidth]{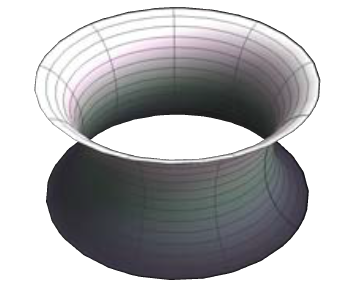}
    \caption{The three pseudospheres: (left) ELL, (center) BEL and (right) HYP. Locally they have the same geometry, $K=-1/r^2$; they are distinguished by global (topological) properties. In the correspondence of Section~\ref{sec:BTZ_spectrum} they are the time slices of the BTZ solution for $M<0$, $M=0$ and $M>0$, respectively.}
    \label{fig:ELLBELHYP}
\end{figure}

The three pseudospheres are related by limits, but at finite values of the parameters they are not isometric to one another: as seen in Section~\ref{sec:EllParHyp}, they are distinguished by global invariants (cone angle, cusp, closed geodesic), even though they are locally isometric, see Fig.~\ref{fig:ELLBELHYP}.
\begin{itemize}
  \item For $c/r = \sin\theta\to0$, ELL tends to BEL: after a shift of the meridian coordinate, $c\sinh(u/r)\to c' e^{u/r}$, the tip moves to infinity and becomes a cusp, the cone angle $2\pi\sin\theta$ goes to zero, and the area $2\pi r^2(1-\sin\theta)$ tends to the area $2\pi r^2$ of BEL.
  \item For $c/r\to0$, HYP is often said to tend to two BELs joined by their tails (Fig.~\ref{fig:HYP}). This is true only away from the waist\footnote{We thank Valeriia Tynianska for pointing this out to us.}. Measuring the arclength $s$ from one Hilbert horizon, the meridian of HYP is
\begin{equation}\label{HYPvsBEL}
  R(s) = \sqrt{r^2+c^2}\,\cosh\frac{s}{r} - r\sinh\frac{s}{r} = r\,e^{-s/r} + \left(\sqrt{r^2+c^2}-r\right)\cosh\frac{s}{r} \,,
\end{equation}
where the first term is BEL, with its Hilbert horizon at $s=0$. The correction, $\simeq (c^2/2r)\cosh(s/r)$, is negligible near the horizon, but it becomes as large as the BEL term at $s\simeq r\ln(2r/c)$, that is precisely the distance of the waist, $s_w = r\,{\rm arcsinh}(r/c)$. The limit is then only pointwise: at any fixed distance from a Hilbert horizon each half of HYP tends to a BEL, but the waist runs away to infinite distance and is never replaced by a cusp. For any $c>0$, around the waist HYP keeps a neck, a nearly cylindrical tube where $R'\simeq0$: the portion with $R<2c$ has length $2r\,{\rm arccosh}\,2\simeq2.63\,r$, independently of $c$, while for BEL the region $R<2c$ is infinitely long. The waist is a closed geodesic of length $2\pi c$, crossed by the geodesics that go from one half to the other, and each half of HYP is an annulus, $\chi=0$, i.e., a \textit{truncated} BEL rather than a BEL with its cusp (Table~\ref{tab:GBbalance}); also, HYP is finite, with height $\simeq 2r[\ln(2r/c)-0.31]$. Correspondingly, the hyperbolic rotation tends to a parabolic one only as a contraction. The limit of ELL is better behaved: there the tip, carrying $2\pi(1-\sin\theta)$, does become the cusp, carrying $2\pi$, and $\chi=1$ is preserved.
  \item The sign $\pm$ in $z(u)$, Eq.~\eqref{emb}, gives a mirror copy of each surface, and the two copies join at the boundary where $R' = 1$: this is the Hilbert horizon for all three surfaces; ELL too is thus made of two pieces, joined at its Hilbert horizon, $R = r\cos\theta$.
\end{itemize}

The Gauss--Bonnet balance of the three surfaces is summarized in Table~\ref{tab:GBbalance}. The bulk curvature is $-A/r^2$, with $A$ the area. For a surface of revolution $d^2\mu = R(u)\,du\,dv$, hence, integrating Eqs.~\eqref{ELL}--\eqref{HYP} over their domains,
\begin{equation}\label{AreasPseudo}
  A = 2\pi\int R(u)\,du = \left\{
  \begin{array}{ll}
    2\pi r^2(1-\sin\theta) & \quad \text{(ELL)} \\
    2\pi r^2 & \quad \text{(BEL)} \\
    4\pi r^2 & \quad \text{(HYP)}
  \end{array}\right.
\end{equation}
and the areas of BEL and HYP do not depend on $c$. For a parallel, the geodesic curvature is $k_g = R'/R$, so that the Hilbert horizon, where $R'=1$, always contributes $+2\pi$, and the waist of HYP, where $R'=0$, contributes nothing (it is a geodesic). The tip of ELL is a conical point carrying the curvature $2\pi(1-\sin\theta)$, exactly opposite to the bulk curvature: with its tip, ELL is balanced like a flat disk. The cusp of BEL, a point at infinity, carries $2\pi$. Removing the cusp turns BEL into an annulus, $\chi = 0$, the missing curvature reappearing as the boundary term of a small circle around it. In what follows, ELL is always considered with its tip.

\begin{table}[htbp]
\centering
\tbl{Balanced Gauss--Bonnet for the three pseudospheres: bulk curvature, contribution of the tip (ELL) or cusp (BEL), boundary terms, and Euler characteristic. ELL is parametrized by $c = r\sin\theta$, HYP by $c>0$; Hh stands for Hilbert horizon.\label{tab:GBbalance}}
{\begin{tabular}{|c|c|c|c|c|}
    \hline
    surface & bulk $\int K\,d^2\mu$ & tip / cusp & boundaries $\oint k_g\,d\ell$ & $\chi$ \\ \hline
    ELL, with tip    & $-2\pi(1-\sin\theta)$ & $+2\pi(1-\sin\theta)$ & $+2\pi$ (Hh) & 1 \\
    BEL, with cusp   & $-2\pi$ & $+2\pi$ & $+2\pi$ (Hh) & 1 \\
    BEL, truncated   & $-2\pi$ & --- & $+2\pi$ (Hh), $0$ (cusp) & 0 \\
    HYP              & $-4\pi$ & --- & $+2\pi$, $+2\pi$ (two Hh) & 0 \\ \hline
  \end{tabular}}
\end{table}

For carbon structures, the bulk term must be carried by defects and/or deformed hexagons, in units of $\pi/3$: $6(1-\sin\theta)$ units for ELL, $6$ for BEL, $12$ for HYP, independently of $c$.

\section{Carbon-made ELL}
\label{sec:ELL}

ELL is the pseudosphere whose rotation is elliptic: its tip is a conical point. It is the natural negative-curvature counterpart of the cone, and the first step towards BEL, that is its limit. We first recall the cone and its singularity, then discuss ELL as a $K<0$ cone, and finally construct the arrangement and the discrete symmetries of its defects, extending the construction of Ref.~\protect\refcite{LobLab_2016}.

\subsection{Cones and the $K<0$ cone}\label{sec:nanocones}

\begin{figure}[htbp]
  \centering
  \begin{minipage}{0.38\textwidth}\centering\includegraphics[width=\linewidth]{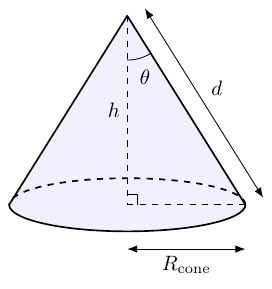}\end{minipage}\hfill
  \begin{minipage}{0.52\textwidth}\centering\includegraphics[width=\linewidth]{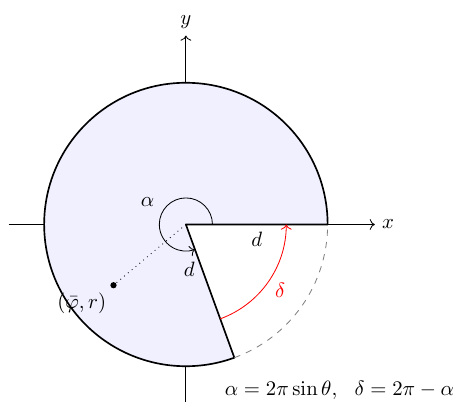}\end{minipage}
  \caption{(Left) The cone in $\mathbb{R}^3$, with half-aperture $\theta$, generatrix $d$, height $h=d\cos\theta$, base radius $R_{\rm cone}=d\sin\theta$. (Right) The same cone cut open and laid flat, a disk of radius $d$ with a missing wedge of angle $\delta = 2\pi-\alpha$, the deficit angle.}
  \label{cone_3d}
\end{figure}

A cone of half-aperture $\theta$ and generatrix $d$ (Fig.~\ref{cone_3d}, left) is a developable surface: cut open and laid flat, it is a disk of radius $d$ with a missing wedge (Fig.~\ref{cone_3d}, right). Equating the base circumference, $2\pi d\sin\theta$, with the length of the arc, $d\,\alpha$, gives $\alpha = 2\pi\sin\theta$, hence the deficit angle
\begin{equation}\label{defang}
  \delta \equiv 2 \pi -  \alpha = 2 \pi (1 - \sin \theta) \,.
\end{equation}
With the line element $dl^2 = d\rho^2 + \rho^2 \sin^2 \theta\, d \varphi^2$, $\varphi\in[0,2\pi]$, the Gaussian curvature is zero everywhere except at the tip,
\begin{equation}\label{GaussCone}
  K (\rho, \varphi) = \frac{\delta}{2 \pi}\, \frac{\delta^{(1)} (\rho)}{\rho \sin \theta} \,, \qquad \int_{cone} K\, d^2 \mu = \delta \,.
\end{equation}
This is a \emph{quasi-regular singularity}: a finite ``curvature charge'' sits at a single point, all curvature invariants stay bounded for $\rho>0$, and radial geodesics reach the tip in finite affine parameter, with no unique continuation.

ELL's line element is $dl^2 = du^2 + (r\sin\theta)^2\sinh^2(u/r)\,dv^2$, with $K=-1/r^2$ \cite{LobLab_2016}. For $u\to0$ it reduces to $du^2 + \sin^2\theta\,u^2dv^2$: the tip of ELL is a conical point with the same deficit angle $2\pi(1-\sin\theta)$ of the cone tangent to it. Cone and ELL share the conical singularity and the geodesic incompleteness, but differ in the distribution of curvature: zero everywhere, with a Dirac delta at the tip, for the cone; constant and negative everywhere for ELL, whose tip carries the positive curvature $2\pi(1-\sin\theta)$ (Table~\ref{tab:GBbalance}). On a lattice: for the cone, hexagons everywhere and a $(6-n)$-gon at the tip; for ELL, heptagons distributed over the surface, while the tip cannot host a net negative defect (a single heptagon there would be a saddle, spoiling the cone). The precise balance is discussed in Section~\ref{sec:ELLgroup}.

\subsection{First attempts with graphene}

\begin{figure}[htbp]
  \centering
  \includegraphics[width=0.32\linewidth,page=1]{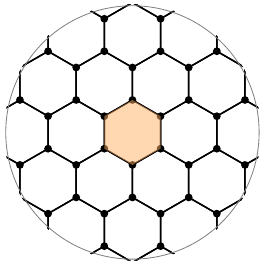}\hfill
  \includegraphics[width=0.32\linewidth,page=2]{fig_volterra}\hfill
  \includegraphics[width=0.32\linewidth,page=3]{fig_volterra}\\[1ex]
  \includegraphics[width=0.32\linewidth,page=4]{fig_volterra}\hfill
  \includegraphics[width=0.32\linewidth,page=5]{fig_volterra}\hfill
  \includegraphics[width=0.32\linewidth,page=6]{fig_volterra}
  \caption{The Volterra process on the hexagonal lattice: removing $n=1,\dots,5$ wedges of $\pi/3$ (shaded), i.e., the deficit angle $\delta_n = n\pi/3$, and gluing the edges, produces cones with $\sin\theta = 1-n/6$ and a $(6-n)$-gon, at the tip, in place of the central hexagon (highlighted in the first panel).}
  \label{Volterra}
\end{figure}

\begin{figure}[htbp]
  \centering
  \includegraphics[width=0.49\linewidth]{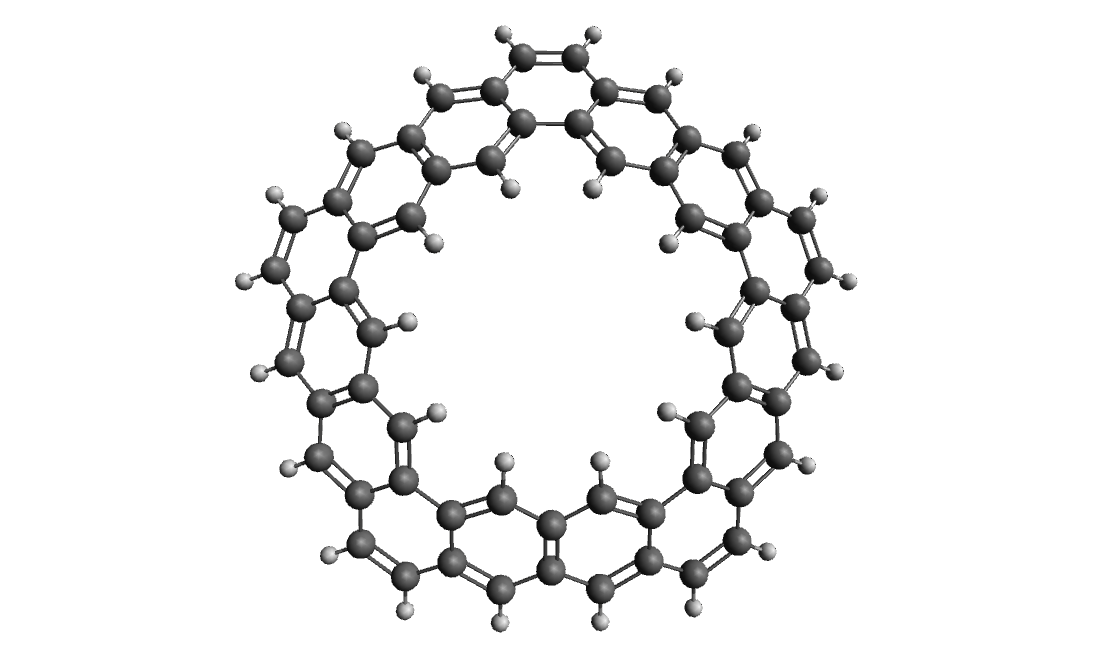}
  \includegraphics[width=0.49\linewidth]{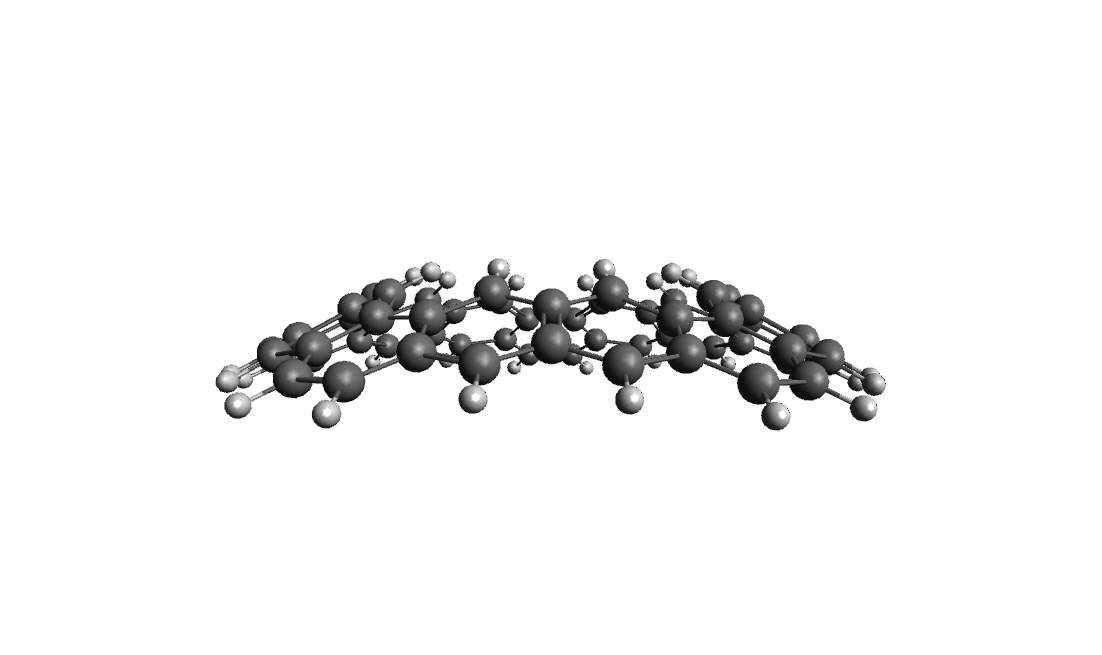}
  \caption{Nanocone with no tip, or a truncated nanocone.}
  \label{fig:NCNnotipX1}
\end{figure}

The first five noticeable $\theta$ angles are those obtained by the Volterra process of cutting wedges of $\pi/3$ alongside the lattice directions and gluing together what is left \cite{Kleinert}, see Fig.~\ref{Volterra}: removing $n$ wedges gives $\sin\theta = 1 - n/6$, the same values that we shall find for the family of ELLs in Section~\ref{sec:ELLgroup}, Eq.~\eqref{ELLk}. A truncated nanocone is shown in Fig.~\ref{fig:NCNnotipX1}.

\subsection{Defects and discrete symmetries of ELL}\label{sec:ELLgroup}

In Ref.~\protect\refcite{LobLab_2016} (see its Supplemental Material) BEL was reached as the limit of a family of ELLs, constructed by accommodating one unit of negative curvature, $-\pi/3$, at a time. We now complete that construction and give the explicit arrangement and symmetries of the defects for the whole family.

\begin{figure}[htbp]
  \centering
  \includegraphics[width=\linewidth]{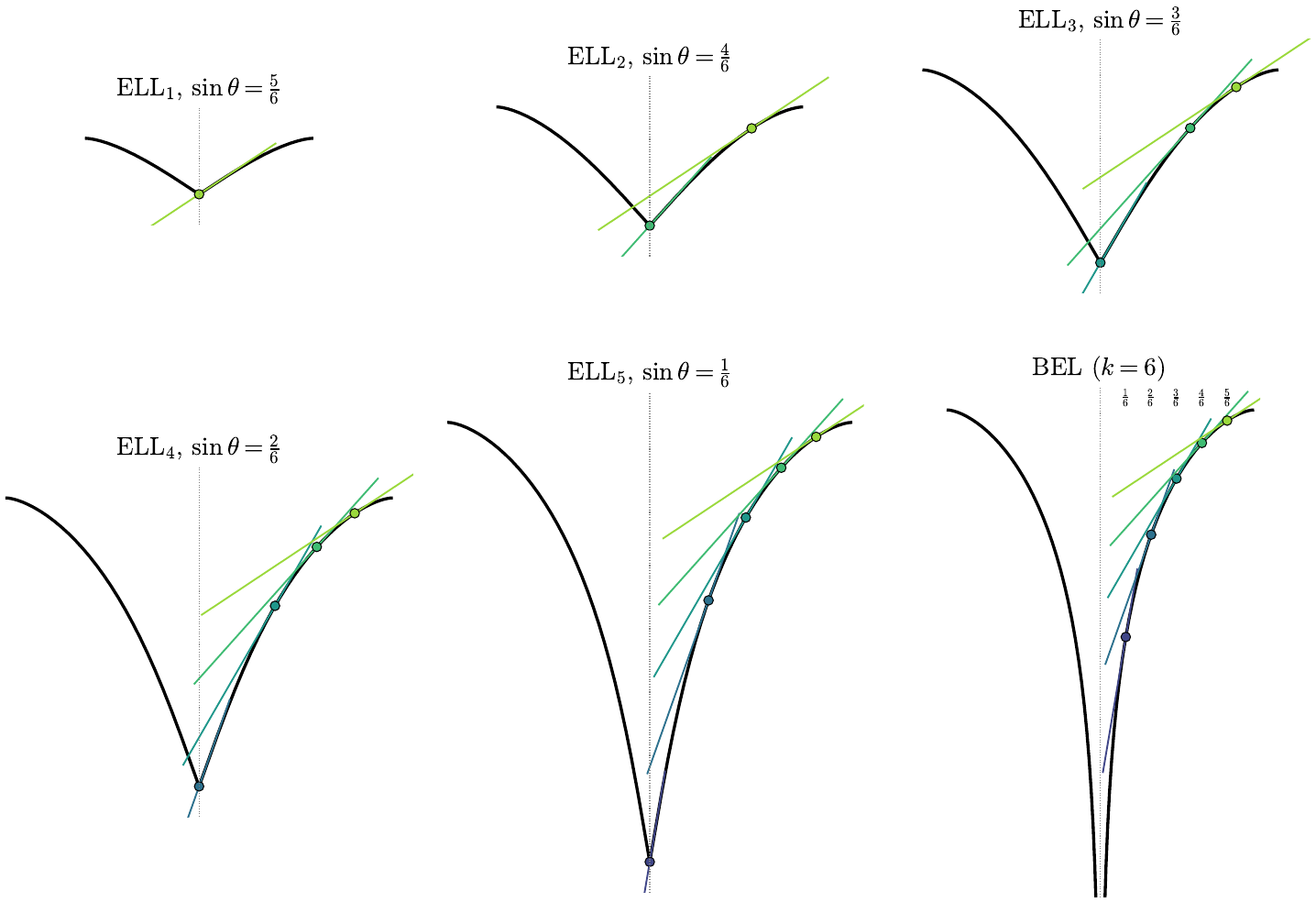}
  \caption{Arrangement of the defects (charge $-1$, curvature $-\pi/3$) on the elliptic pseudospheres ELL$_k$: each defect sits on the parallel where the surface is tangent to the cone of half-aperture $\alpha_n = \arcsin(n/6)$. In the limit $k\to6$ one recovers BEL, with the defects at $R/r = 1/6,\dots,5/6$ (and one at infinity).}
  \label{fig:ELLconstruction}
\end{figure}

The area of ELL is $A = 2\pi r^2(1-\sin\theta)$, Eq.~\eqref{AreasPseudo}, hence its bulk curvature is $-2\pi(1-\sin\theta)$. The member $\mathrm{ELL}_k$ of the family carries $k$ units, $-k\pi/3$, i.e.,
\begin{equation}\label{ELLk}
  \sin\theta_k = \frac{6-k}{6}\,,\qquad k = 1,\dots,5\,,
\end{equation}
with $k=6$ being the limiting case of BEL. For a surface of revolution, $dR/du = \sin\alpha$, with $\alpha$ the angle between the axis and the tangent to the meridian, i.e., the half-aperture of the cone tangent to the surface along that parallel. The construction of Ref.~\protect\refcite{LobLab_2016} places the defects at the parallels where the surface is tangent to the cones of half-aperture
\begin{equation}\label{alphan}
  \sin\alpha_n = \frac{n}{6}\,,
\end{equation}
which are also the parallels where the surface, measured from the tip, has accumulated $n-(6-k)$ units of curvature. For ELL, $R' = \sin\theta\cosh(u/r)$, hence the defects sit at
\begin{equation}\label{ELLdefects}
  \cosh\frac{u_n}{r} = \frac{n}{6-k}\,,\qquad
  R_n = \frac{r}{6}\sqrt{n^2-(6-k)^2}\,,\qquad n = 6-k,\dots,5\,.
\end{equation}
The first value, $n = 6-k$, is the tip, $R=0$, while the value $n=6$, that is not occupied, gives the Hilbert horizon, $R = r\cos\theta_k$: the boundary is the ``sixth cone'', the plane. The $k-1$ defects off the axis are related by the rotations of $\mathbb{Z}_{k-1}$ (an elliptic subgroup, see Section~\ref{sec:EllParHyp}), combined with the discrete steps along the meridian $u_n = r\,{\rm arccosh}[n/(6-k)]$. By analogy with $F_Y$, that is the hyperbolization of the icosahedral group $Y$, we call this discrete structure $F_{C_{k-1}}$, as it is the hyperbolization of the cyclic group $C_{k-1}$:
\begin{equation}\label{FCk}
  F_{C_{k-1}} \equiv \left\{ (u,v) \,\Big|\, v = \frac{2\pi\,[n-(6-k)]}{k-1}\,,\ u = r\,{\rm arccosh}\frac{n}{6-k}\,,\ n = 6-k+1,\dots,5 \right\} \,.
\end{equation}
In the limit $k\to6$, ${\rm arccosh}(n/\epsilon)\simeq\ln(2n/\epsilon)$, and Eq.~\eqref{ELLdefects} gives $R_n = n\,r/6$ and the discrete ``boosts'' $u_n = r\ln(n/6)$ of the structure $F_Y$ of BEL, Eq.~\eqref{FY}, with its $\mathbb{Z}_5$ symmetry. The construction is illustrated in Fig.~\ref{fig:ELLconstruction}.

The Gauss--Bonnet balance of Table~\ref{tab:GBbalance} tells us what the tip must carry. With the tip, the $k$ units of negative bulk curvature are exactly compensated by the cone angle at the tip. If one of the $k$ units of negative curvature is placed at the tip, as in the construction above, the tip carries the net positive charge $(k-1)\pi/3$, compensated by the $k-1$ heptagons off the axis, and the structure is balanced as a flat disk.

\section{Carbon-made BEL}
\label{sec:BEL}

BEL is the case where both the group-theoretical construction and the simulations were carried out \cite{LobLab_2016}; here we recall the main results, in the light of the balanced Gauss--Bonnet theorem.

\textit{Topology and number of defects.} The area of BEL is $2\pi r^2$, Eq.~\eqref{AreasPseudo}, hence its bulk curvature is $-2\pi$, i.e., six units of $-\pi/3$: six heptagons for an ideal carbon BEL. One of them sits at the cusp, at infinity, and is removed in any practical (truncated) realization, so that five heptagons are expected on the surface (Table~\ref{tab:GBbalance}).

\textit{The discrete group.} The six defects are the vertices of a ``Lobachevsky Beltrami polyhedron'', constructed by deforming the icosahedral group $Y = (C_2,C_3,C_5)$, oriented with one vertex at the south pole, mapped to the tip of BEL. The $C_5$ symmetry along the parallels is preserved, while the $C_5$ symmetry along the meridians is ``hyperbolized'' into a discrete boost, giving \cite{LobLab_2016}
\begin{equation}\label{FY}
  F_Y \equiv \left\{ (u,v) \,\Big|\, v = \frac{2\pi n}{5}\,,\ u = r\ln\frac{|n|}{6}\,,\ n\in\mathbb{Z}_5 \right\} \,,
\end{equation}
where $n=0$ is the cusp, at infinity, and the angles are taken modulo $2\pi$ \cite{LobLab_2016}. Strictly speaking, $F_Y$ is not a group but a finite set of points, the positions of the defects: its angular part is generated by the cyclic group $\mathbb{Z}_5$ of the rotations $v\mapsto v+2\pi/5$, while the steps along the meridian, $u_n = r\ln(n/6)$, are fixed by the curvature condition and are not uniform, so that the set is not closed under composition. We keep the name $F_Y$ of Ref.~\protect\refcite{LobLab_2016}, and use analogous names below, for these discrete structures, the ``hyperbolized'' counterparts of the corresponding finite groups; the genuine discrete groups behind them are discussed in Section~\ref{sec:Summary}. The sign of $n$ gives the chirality of the helix on which the defects sit. The defects are at the radii $R_n/r = n/6$, $n=1,\dots,5$, where BEL is tangent to the cones of half-aperture $\arcsin(n/6)$, and where the surface has accumulated $n$ units of curvature from the cusp, see Section~\ref{sec:ELLgroup}.

\textit{Simulations.} In Ref.~\protect\refcite{LobLab_2016} carbon BELs with 1146, 2146 and 5506 atoms were obtained by a triangulation of the surface with a Lennard-Jones potential (minimization with the fast inertial relaxation engine, FIRE), followed by Voronoi dualization, relaxation with a Stillinger--Weber potential and density-functional tight-binding calculations. The extrinsic curvature produces a diverging stretching at the Hilbert horizon and a contraction at the tip; the first is cured by pinning the atoms at the horizon, and by adding a flat ring of graphene beyond it. Heptagons and pentagons form charged and neutral scars, distributed along a helix; the number of charged scars converges to the expected five as the size grows, in agreement with $F_Y$. A large hexagonal circulation tangent to the Hilbert horizon appears only when no charged scar crosses the horizon, and it is the signature that a BEL has been obtained. The 1146-atom BEL is stable at the density functional theory (DFT) and density-functional tight-binding (DFTB) levels also when the constraint to the surface is released. In the language of this paper: the bulk charge is carried by five charged scars, the boundary term is controlled by the pinned Hilbert horizon and the flat ring, and only then the structure is a BEL.

\begin{figure}[htbp]
  \centering
  \includegraphics[width=0.75\linewidth]{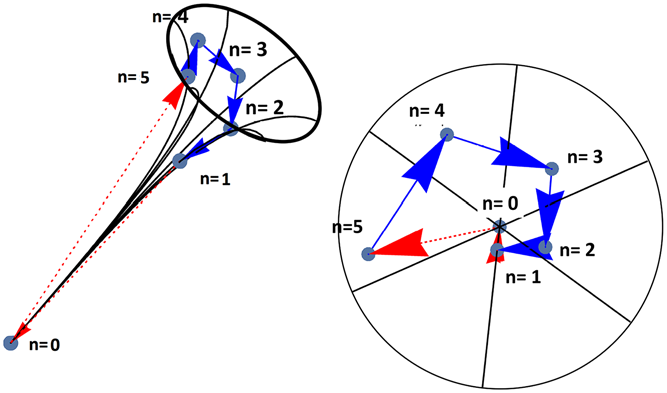}\\[1ex]
  \includegraphics[width=0.95\linewidth]{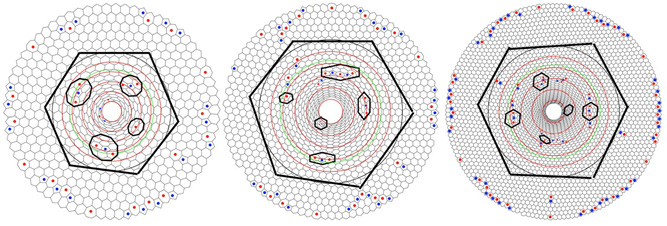}
  \caption{(Top) Action of $F_Y$ on the vertices of the Lobachevsky Beltrami polyhedron, side and top views; the vertex at infinity ($n=0$) is identified with $n=5$. (Bottom) Optimized carbon BELs with 1146, 2146 and 5506 atoms (top views), with the charged scars and the hexagonal circulation tangent to the Hilbert horizon. Reproduced from Ref.~\protect\refcite{LobLab_2016}.}
  \label{fig:BELsummary}
\end{figure}

\section{Carbon-made HYP}
\label{sec:HYP}

HYP is the pseudosphere whose rotation is hyperbolic, and the only one with two Hilbert horizons and a closed geodesic, the waist. Its bulk curvature is $-4\pi$, i.e., twelve units of $-\pi/3$, for any value of $c$ (Table~\ref{tab:GBbalance}), and each half, between the waist and one Hilbert horizon, carries six units, like a BEL. We discuss two complementary routes to carbon HYPs: \textit{by design}, as polycyclic aromatic hydrocarbons with an ideal arrangement of heptagons, and \textit{by simulated growth}, through nano-extrusion and annealing \cite{klaver2025stability}. The first route places the twelve units of curvature in twelve isolated heptagons, symmetrically arranged; the second lets the carbon find its own arrangement. That the two lead to the same surface, and under which conditions, is a direct test of the balanced Gauss--Bonnet theorem.

\subsection{Ideal HYPs by design}\label{sec:HYPdesign}

\begin{figure}[htbp]
  \centering
  \subfloat[hexagonal]{\includegraphics[width=0.32\linewidth]{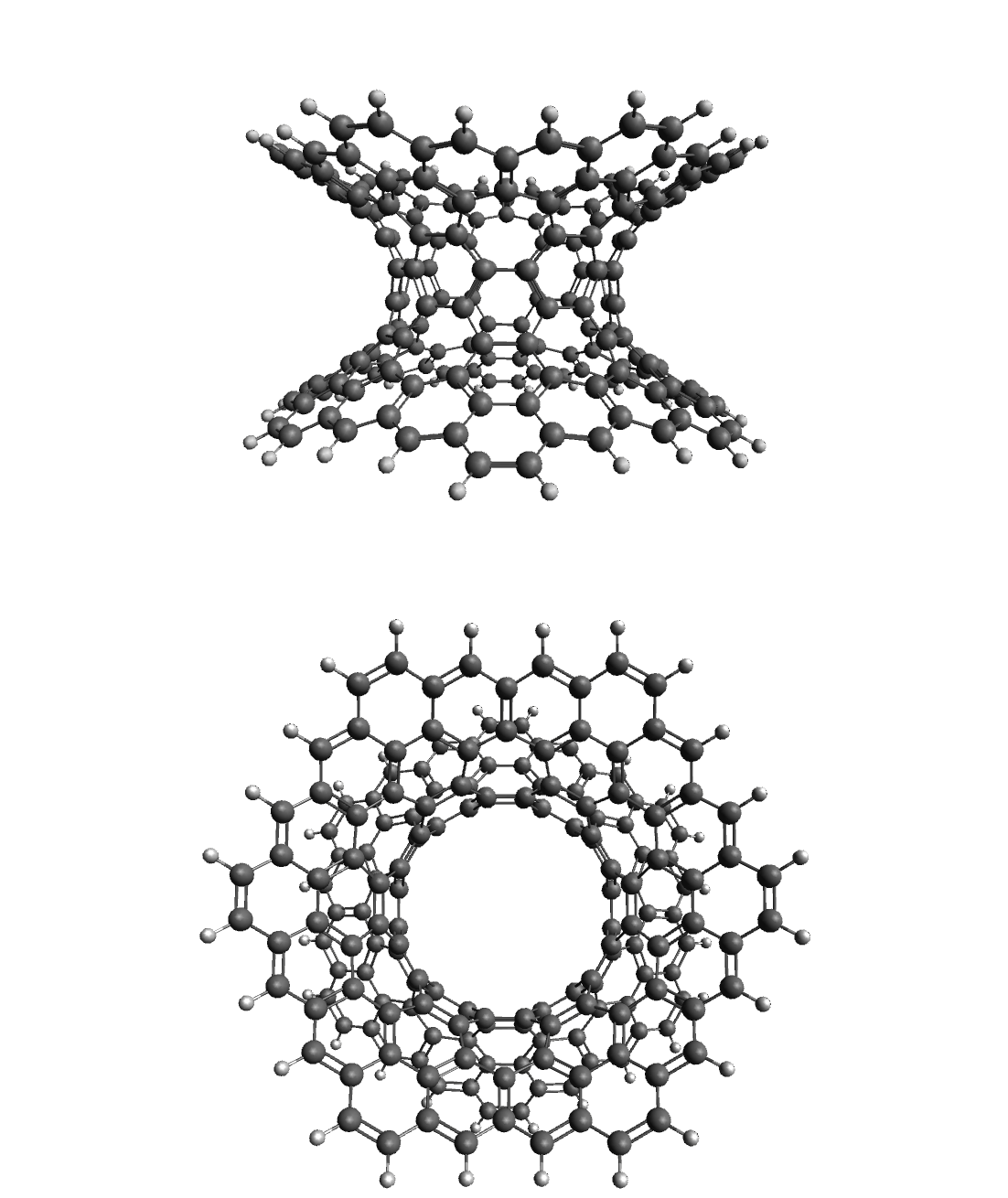}\label{hyp11-hex}}
  \subfloat[trigonal]{\includegraphics[width=0.32\linewidth]{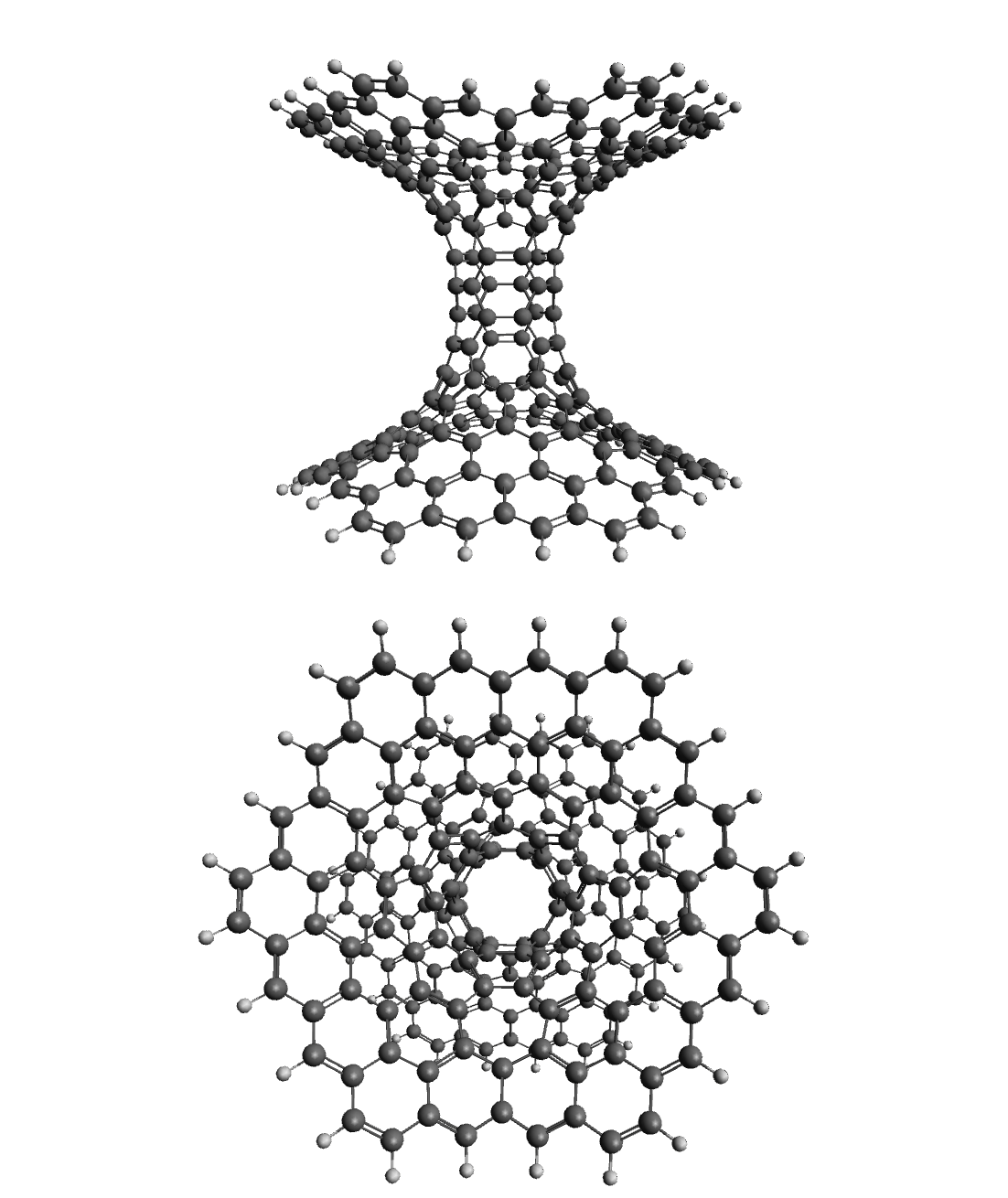}\label{hyp11-tri}}
  \subfloat[improper 4-fold]{\includegraphics[width=0.32\linewidth]{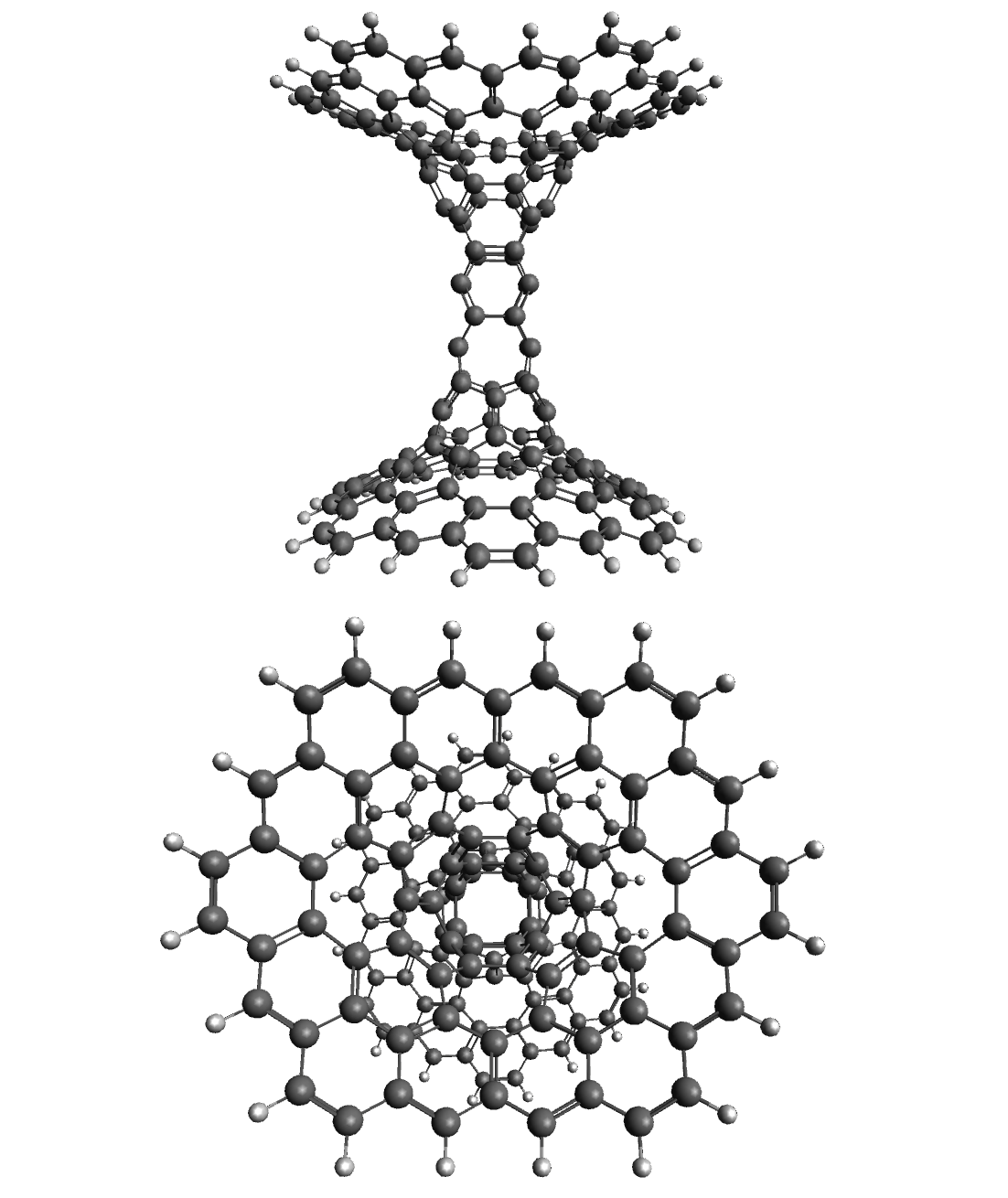}\label{hyp11-imp}}
  \caption{Carbon-made HYP(1,1)s, with heptagonal rings surrounded by hexagons: (a) hexagonal, (b) trigonal, and (c) improper 4-fold symmetrical.}
  \label{fig:hyp1-1}
\end{figure}

A HYP, being finite, can be realized as a single polycyclic aromatic hydrocarbon (PAH) containing heptagonal rings. Compared to fullerenes, which only contain carbon atoms and map to closed surfaces, PAHs possess a boundary where the hydrogen atoms are located. Moreover, since each heptagon carries a fixed amount of Gaussian curvature, only a homogeneous distribution of these---as in a heptagonal lattice---can realize an ideal surface with constant Gaussian curvature. As with Goldberg polyhedra, the arrangement of heptagons can be represented by indices $(i,j)$ along the two lattice directions. The smallest family, HYP(1,0), has members made of heptagons only. It contains three molecules: \ce{C48H24} ($D_{6h}$), \ce{C48H24} ($D_{3h}$) and \ce{C44H20} ($D_{2d}$). In the family HYP(1,1), the negatively curved counterpart of \ce{C60}, each heptagon is surrounded by seven hexagons. The atoms can in principle still lie on the mathematical surface: \ce{C192H48}, \ce{C228H48} and \ce{C168H40}, with the same three symmetries (Fig.~\ref{fig:hyp1-1}). Larger structures, such as the chiral trigonal HYP(2,1) with 384 carbon atoms, are obtained by spreading the heptagons further apart. All these molecules contain exactly twelve heptagonal rings, as the total curvature, $-4\pi$, requires. As for large fullerenes containing isolated defects, the larger the structure, the flatter the areas away from the defects will appear. In the limit, these PAHs would approach triangulated polyhedra (e.g., the lateral surface of a hexagonal antiprism for a HYP with a hexagonal symmetry). The curvature would be concentrated at the heptagons in the vertices of the polyhedra while the hexagons would both cover their faces and easily bend over their edges.

Which arrangement of defects is preferred? Total energies of small PAHs containing only 5-, 6- and 7-membered rings, computed with the self-consistent-charge density-functional tight-binding method, show that heptagons are energetically more expensive than pentagons, and that the two signs of curvature behave differently: for $K>0$ a 5-7-5 scar is more stable than an isolated pentagon, while for $K<0$ an isolated heptagon is more stable than a 7-5-7 scar. This suggests a principle of \emph{curvature matching}: to approach the ground state, the curvature produced by a local tiling must match the curvature that the surface would form there; scars, as discussed in Ref.~\protect\refcite{Bowick:2000aa}, split and dilute the curvature of isolated defects. For HYP, it points to isolated heptagons, homogeneously distributed, as in the ideal molecules above.

On the experimental side, small molecules with one ([7]circulene) and two ([7.7]circulene) heptagonal rings have been synthesized \cite{Yamamoto:1983aa,Yamamoto:1991aa}. Heptagons can be formed by Scholl reactions of macrocyclic precursors \cite{Cheung:2023aa}, but each additional ring requires its own precursors and catalysts, and the strain grows with the number of heptagons. Larger molecules require multi-step processes, in which monomers are assembled through cross-coupling reactions (Suzuki--Miyaura, Yamamoto, Scholl); recently, a molecule with four heptagonal rings, a fragment of a Schwarzite, has been synthesized \cite{Spengler:2026aa}. Monomers for the trigonal HYP(2,0) can be designed along these lines.

\subsection{A proposal for the discrete symmetries of HYP}\label{sec:HYPgroup}

Applying to each half of HYP the rule found for ELL and BEL, i.e., placing the defects on the parallels where the surface is tangent to the cones with $\sin\alpha_n = n/6$, and using $R' = (c/r)\sinh(u/r)$, one finds the radii
\begin{equation}\label{HYPdefects}
  R_n = \sqrt{c^2 + \left(\frac{n\,r}{6}\right)^2}\,,\qquad n = 0,\dots,5\,,
\end{equation}
on each half, as for BEL in Ref.~\protect\refcite{LobLab_2016}: $n=0$ is the waist (the counterpart of the cusp of BEL), while the value $n=6$, that is not occupied, gives the Hilbert horizon, $R_6=\sqrt{c^2+r^2}$. In the limit $c\to0$, for $n\ge1$, these tend to the radii $n r/6$ of BEL, while the defect $n=0$ stays on the waist (see the discussion of Section~\ref{sec:PseudoLimits}). The rotations are now hyperbolic, and the reflection $u\mapsto-u$ exchanges the two halves. On each half the six defects are related by the rotations of $C_6$, combined with discrete steps along the meridian, $u_n = r\,{\rm arcsinh}(n r/6c)$, and the two halves are exchanged by a twofold rotation: by analogy with $F_Y$ and $F_{C_{k-1}}$, we call this discrete structure $F_{D_6}$, the hyperbolization of the dihedral group $D_6$. In the limit $c\to0$ the defect on the waist ($n=0$) of each half runs away to infinite distance, and, away from the waist, each half approaches the BEL of $F_Y$. This arrangement, one defect per level along a helix, is not the only one compatible with the integrated curvature: the ideal molecules of Section~\ref{sec:HYPdesign}, with symmetries $D_{6h}$, $D_{3h}$ and $D_{2d}$, place six, three or two heptagons on the same level, just as the icosahedron, along a fivefold axis, has the structure $1+5+5+1$. Which of these arrangements corresponds to the minimal-energy carbon HYP, for given $r$ and $c$, is an open question.

\subsection{HYP by nano-extrusion and annealing}\label{sec:HYPextrusion}

\begin{figure}[htbp]
  \centering
  \includegraphics[width=0.95\linewidth]{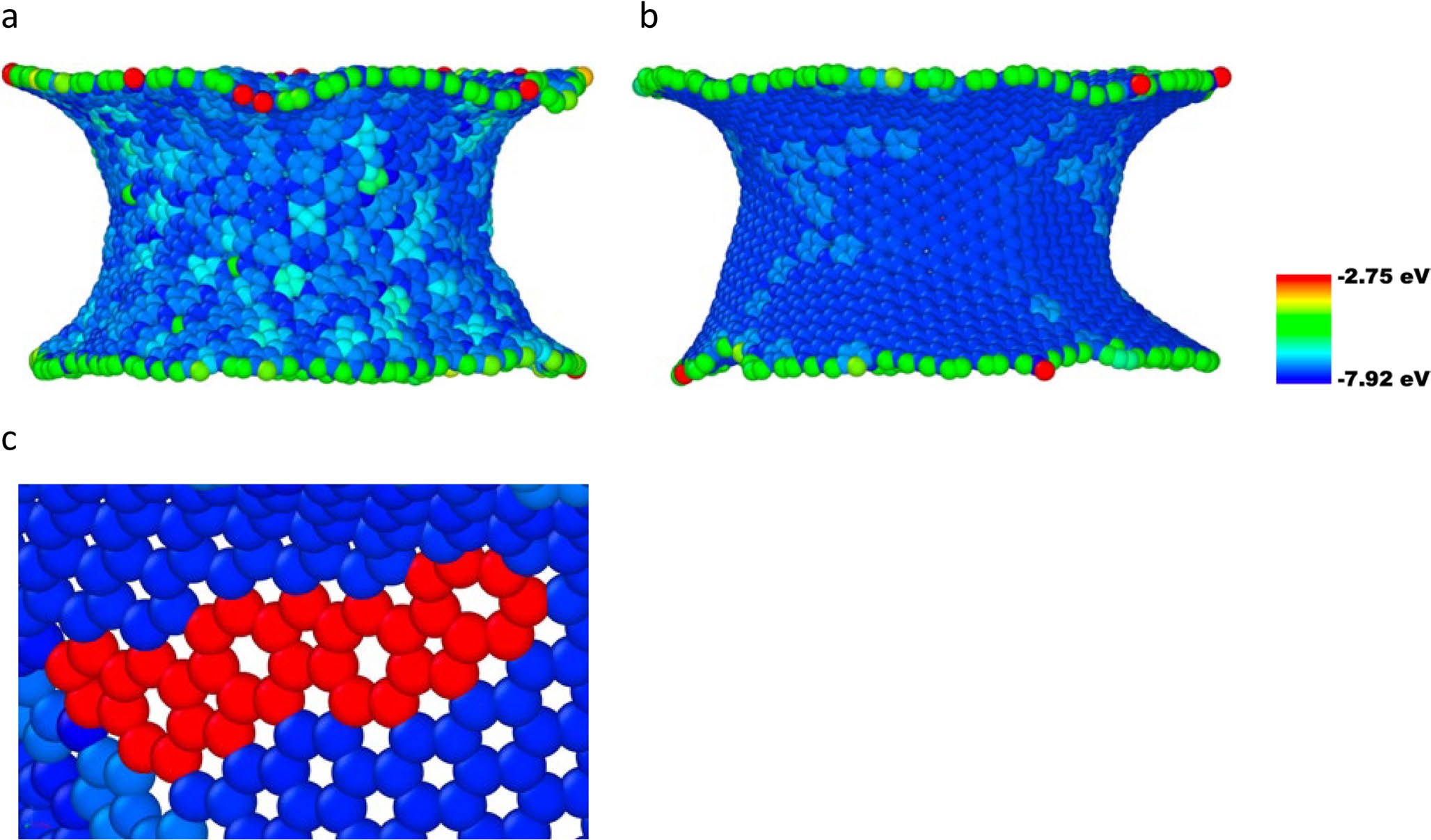}
  \caption{Carbon HYP obtained by nano-extrusion and annealing, with free edges: (a) before and (b) after annealing at $3500$\,K, atoms colored by potential energy; (c) a heptagon-pentagon grain boundary. Reproduced from Ref.~\protect\refcite{klaver2025stability}.}
  \label{fig:HYPextrusionFree}
\end{figure}

A completely different route was followed in Ref.~\protect\refcite{klaver2025stability}, using molecular dynamics (with the LAMMPS code and the second-generation reactive empirical bond-order potential, REBO2). Carbon atoms are first forced, by a simulated extrusion press, down a thin volume shaped as a HYP ($r = c$, axis length $28.5$\,\AA, minimum and maximum diameters $40$ and $56.5$\,\AA), bounded by a strong Lennard-Jones wall and a repulsive harmonic wall. The extruded carbon has the areal density of graphene, but it is an unrealistic, highly defective and unstable precursor. It is then annealed, still constrained to the HYP shape, at temperatures up to $3500$\,K and for times up to $0.1\,\mu$s: most defects annihilate in pairs, or migrate to the free edges, and the structure turns into polycrystalline, curved graphene, with grain boundaries made mostly of alternating heptagons and pentagons (Fig.~\ref{fig:HYPextrusionFree}). During the annealing, the total Gaussian curvature fluctuates and develops towards the expected $K_{tot} = -4\pi$.

The resulting arrangement of defects is very far from the ideal one of Section~\ref{sec:HYPdesign}: in the structure with free edges there are more than fifty heptagon-pentagon pairs, a few isolated heptagons in excess, and a complex quad-vacancy cluster of three octagons and two pentagons, with a defect density about one order of magnitude higher than the minimal one. Nonetheless, the detailed counting of all the non-hexagonal rings, of the triangular rings, and of the $sp$ and $sp^3$ carbons, reported in Appendix~\ref{app:CountingHYP}, gives exactly twelve units of negative curvature, i.e., $-4\pi$. The counting was carried out on one of the HYPs embedded between the two periodic sheets, and some of the scars extend across the Hilbert horizons into the sheets. For this closed structure (two tori joined by a neck, $\chi=-2$) the total, $-4\pi$, is fixed by topology; the non-trivial information is where the charge sits: on the HYP and across its junctions with the sheets, which are otherwise flat. The bulk term of the Gauss--Bonnet balance is thus fully accounted for by the defects, but distributed in a completely different way. Such a defect-rich curved graphene is interesting also beyond geometry: the magnetic moments carried by its defects, and by those of its hydrogenated counterpart, graphane, have been studied in Ref.~\protect\refcite{KlaverLegut2026magnetic}.

\begin{figure}[htbp]
  \centering
  \includegraphics[width=0.95\linewidth]{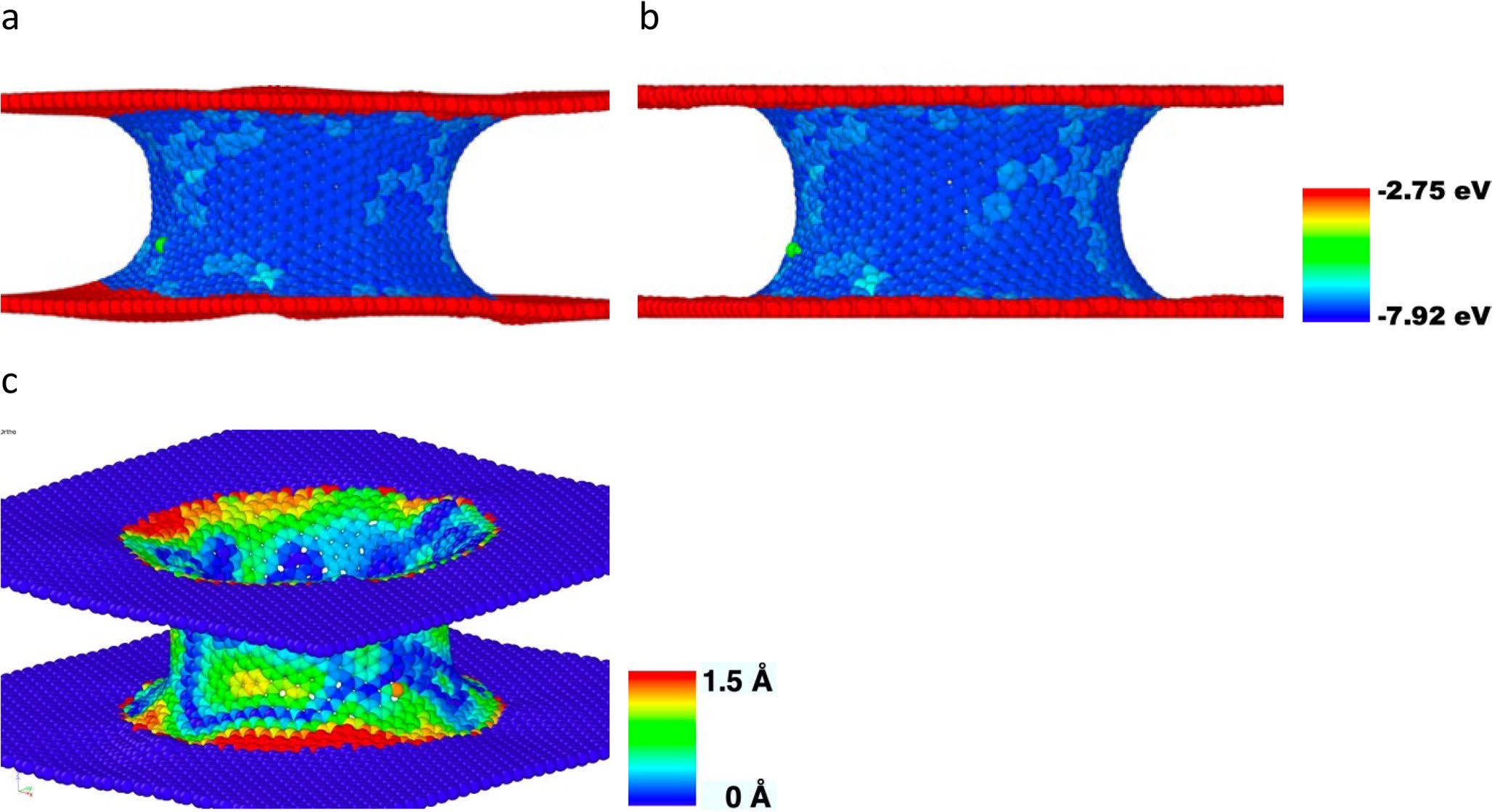}
  \caption{Annealed and relaxed carbon HYP embedded in flat periodic graphene sheets: (a) without and (b) with flat walls on the sheets, atoms colored by potential energy; (c) deviation of the atoms from the mathematical HYP. Reproduced from Ref.~\protect\refcite{klaver2025stability}.}
  \label{fig:HYPextrusion}
\end{figure}

What decides the shape is then the boundary term. When the structure is released, with free edges, it deviates considerably from the mathematical HYP. When the free edges are removed, by embedding the HYP between two periodic flat graphene sheets, so that the Hilbert horizons are no longer material edges (Fig.~\ref{fig:HYPextrusion}), the carbon atoms sit on average $0.6$\,\AA\ from the mathematical surface, and only $0.2$\,\AA\ when the sheets are kept flat, as if they were stiff stacks of many sheets. The structure also proved stable against $20\%$ elongation, $20^\circ$ shearing and annealing at $1500$\,K. This is the balanced Gauss--Bonnet theorem at work: the same total curvature charge, $-4\pi$, gives the same surface, whether carried by twelve isolated heptagons or by tens of scars, provided that the boundary term is the right one.

\section{Applications: analogue gravity and desalination}
\label{sec:Applications}

The applications of these structures can be many. Carbon is used today in a wide range of contexts, from electronics \cite{PacoReview2009} to energy storage \cite{SimonGogotsi2008} and filtration \cite{GeimScience2014}, and curved carbon structures are certainly at work in many of them in ways that are not yet recognized. Much of what we discussed is also of purely mathematical interest: the surfaces of constant negative curvature \cite{eisenhart1909treatise,ovchinnikov} and their link with the solitons of the sine-Gordon equation \cite{mclachlan,RogersSchief2002}, the discrete geometry of deformed lattices \cite{vincenzino}, the Lobachevsky geometry and its discrete groups \cite{Katok1992,Magnus1974,Boettcher2022Crystallography}. Among all these possible directions, we focus here on the two that we have been pursuing: the realization of the BTZ black hole and wormhole in the laboratory, and water desalination.

\subsection{The BTZ black-hole spectrum and wormholes}
\label{sec:BTZ_spectrum}

We briefly recall here how the three pseudospheres emerge from the spectrum of the BTZ black hole when the time is frozen (time-slice approach). The differences between the pseudospheres proper and the structures obtained from the limits of HYP, together with the classical traversability of the corresponding carbon BTZ wormhole, are discussed in Ref.~\protect\refcite{iorio2026classical}.

\begin{figure}[h]
  \centering
  \includegraphics[width=0.8\textwidth]{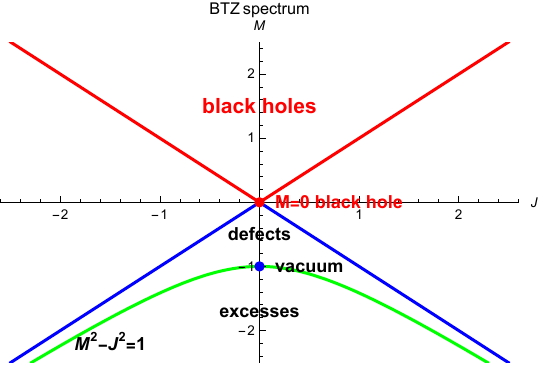}
  \caption{The spectrum of the BTZ solutions in the $(J,M)$ plane, with $l=1$ (see Ref.~\protect\refcite{ZanelliBTZspectrum2019}): black holes ($M\ge|J|$), conical defects ($-1<M<0$), the anti-de Sitter vacuum ($M=-1$, $J=0$) and conical excesses ($M<-1$). The regions $|J|>|M|$, left blank, also correspond to naked singularities.}
  \label{fig:BTZspectrum}
\end{figure}

The BTZ metric \cite{BTZ1992} for a non-spinning black hole is
\begin{equation}\label{BTZ}
ds_{BTZ}^2=\left(\frac{\mathcal{R}^2}{l^2}-M\right)d\mathcal{T}^2- \left(\frac{\mathcal{R}^2}{l^2}-M\right)^{-1} d\mathcal{R}^2 -\mathcal{R}^2 d{\mathcal{V}}^2,
\end{equation}
with $\Lambda = -l^{-2}$ and horizon at $\mathcal{R}_{hor} = l\sqrt{M}$. At fixed $\mathcal{T}$, the spatial part is $SO(2)$-symmetric \cite{BHTZ}, and identifying it with the line element of a surface of revolution, $du^2 + R^2(u)\,dv^2$, gives $v = \mathcal{V}$ and $R(u) = \mathcal{R}$: the radial coordinate of the carbon structure coincides with that of the BTZ space (in the Weyl approach, instead, the two differ, and the BTZ horizon is mapped to the Hilbert horizon of BEL \cite{iorio2014}). The meridian part then gives
\begin{equation}\label{defining_R}
\frac{dR}{du} = \frac{1}{l} {\sqrt{R^2-l^2M}} \,,
\end{equation}
whose solutions, for the three signs of $M$ in the BTZ spectrum \cite{ZanelliBTZspectrum2019}, are
\begin{align}
& M<0: \quad R(u)=l\sqrt{|M|}\sinh(u/l) \,, \label{BTZ_ELL} \\
& M=0: \quad R(u)= \alpha \, e^{u/l} \,, \\
& M>0: \quad R(u)=l\sqrt{M}\cosh(u/l) \,.
\end{align}
Comparing with Eqs.~\eqref{ELL}--\eqref{HYP}, each pseudosphere realizes one part of the BTZ spectrum: ELL for $M<0$ (no horizon), BEL for $M=0$ (horizon at the cusp, at spatial infinity), HYP for $M>0$, the only one with the horizon within the surface, at the waist, $R = c = R_{hor}$ (Fig.~\ref{fig:ELLBELHYP}). This result is implicit in earlier investigations of the BTZ spectrum \cite{brill20002+,brill2000black,brill1996geometry}; here we make it explicit. The parameters are identified as
\begin{equation} \label{parameterident}
    r = l\,, \qquad c = r \sqrt{M} \quad (\mbox{not for BEL})\,, \qquad {\rm i.e.,} \qquad \Lambda = - \frac{1}{r^2} \,, \quad M =  \frac{c^2}{r^2} \,.
\end{equation}
The whole spectrum \cite{ZanelliBTZspectrum2019} is sketched in Fig.~\ref{fig:BTZspectrum}.

\subsection{Desalination}
\label{sec:Desalination}

Purification and desalination of seawater is a global issue of vital strategic importance, given the growing scarcity of drinking water caused by climate change.

Among the innumerable applications of graphene, and of other allotropic forms of carbon, such as nanotubes and nanocones, its use as a ``molecular sieve'' \cite{GeimScience2014}, for filtering polluted, fresh or marine waters, to obtain clean or drinkable water, is by now an established reality, as demonstrated by a large literature \cite{Memisoglu2023GrapheneNanocomposite,Lazarenko2022RecentAdvances,Seo2018AntifoulingGraphene,Boretti2018OutlookGraphene,Hummer2001WaterConduction,Koga2001OrderedIceCNT,Wang2020UltrathinCNT,Das2014CNTBrightFuture,GeimScience2014,Buelke2018GrapheneOxideReview}, see also the review \cite{Castelletto2021}. To this end, a very important role is played by the geometry of these carbon structures. A case of great interest is that of the theoretical study, computer simulation and laboratory realization of ``hourglass''-shaped carbon structures, studied, e.g., in Ref.~\protect\refcite{hourglass2021}.

The rationale, behind those studies, is to create filtering structures, bio-inspired by aquaporins in biological organisms \cite{Gravelle2013HourglassAquaporins}. The latter are the so-called ``channel proteins'' which, also by virtue of their hourglass shape, facilitate the rapid passage of water, inside or outside the cells of specific tissues, while effectively filtering other molecules which are not water, as sodium ions or potassium.

\begin{figure}[htbp]
  \centering
  \includegraphics[width=\linewidth]{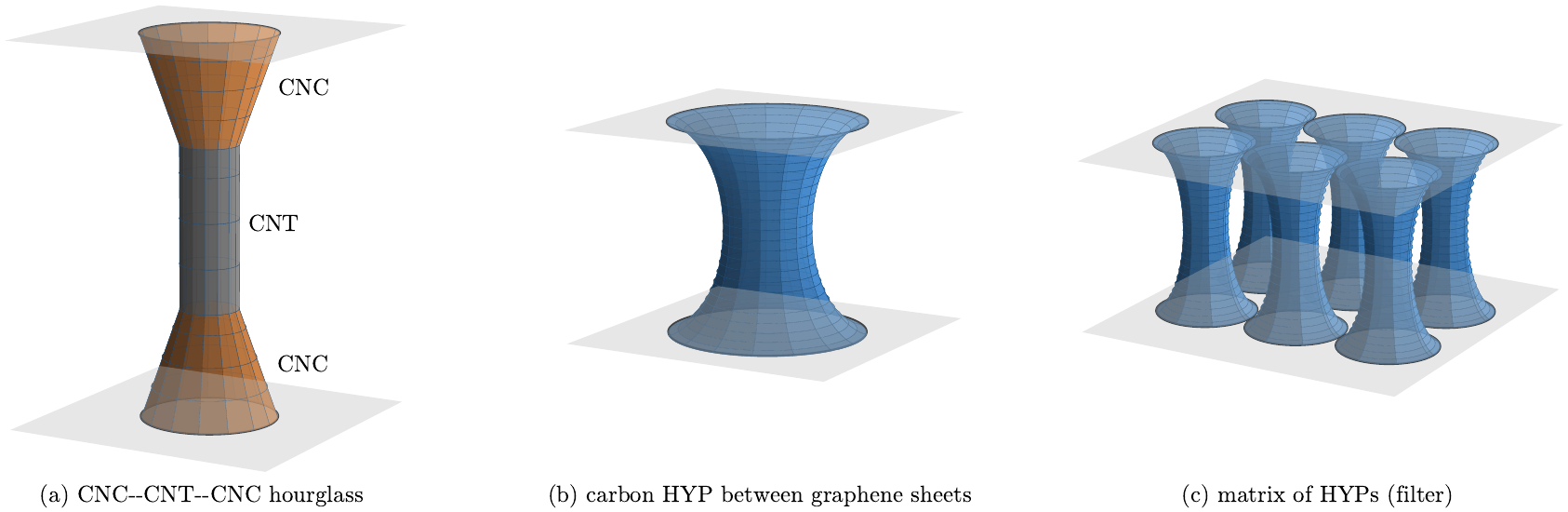}
  \caption{(a) The hourglass-shaped carbon nanopore of Ref.~\protect\refcite{hourglass2021}: two carbon nanocones (CNC) joined to a carbon nanotube (CNT), between two graphene sheets (here with half-aperture $\alpha = 19.45^\circ$). (b) The carbon HYP between two graphene sheets proposed in Ref.~\protect\refcite{Iorio2024DesalinationPseudospheres}, with the same total curvature, $-4\pi$, distributed over the whole surface. (c) A matrix of carbon HYPs between two graphene sheets, acting as a filter \cite{Iorio2024DesalinationPseudospheres}.}
  \label{fig:desalination}
\end{figure}

The hourglass-shaped carbon nanopores studied in Ref.~\protect\refcite{hourglass2021} consist of two carbon nanocones (CNCs) connected, by their truncated tips, to the two ends of a carbon nanotube (CNT), and bounded by two parallel graphene sheets, see Fig.~\ref{fig:desalination}(a). With molecular dynamics simulations of a $0.5$\,M NaCl solution, nanotubes of chirality $(6,6)$, $(8,8)$ and $(10,10)$ and nanocones of half-aperture $\alpha = 41.8^\circ, 30.0^\circ, 19.45^\circ, 9.6^\circ$ (and $0^\circ$, i.e., the bare nanotube), the authors of Ref.~\protect\refcite{hourglass2021} found that the conical entrances enhance the water flux by a factor up to about $4$ with respect to the bare nanotube, the optimal half-aperture being $9.6^\circ$ for all the nanotubes considered, in line with the hydrodynamic predictions of Ref.~\protect\refcite{Gravelle2013HourglassAquaporins}. The narrowest pores, built on the $(6,6)$ nanotube, showed complete salt rejection for all half-apertures, while for the wider ones the salt rejection decreases slightly, remaining above $96\%$.

The half-apertures considered in Ref.~\protect\refcite{hourglass2021} are those that preserve the hexagonal lattice of the nanocones, and they are precisely the angles $\alpha_n = \arcsin(n/6)$, $n = 4,3,2,1$, of Eq.~\eqref{alphan}: the Volterra cones of Fig.~\ref{Volterra}, obtained by removing $6-n$ wedges of $\pi/3$, and the same cones tangent to ELL and BEL at the locations of their defects (Section~\ref{sec:ELLgroup}). In particular, the optimal angle, $9.6^\circ = \arcsin(1/6)$, corresponds to the narrowest of these cones, tangent to BEL at $R = r/6$, the last defect before the cusp.

The balanced Gauss--Bonnet theorem tells us something more about these structures. For a surface of revolution, the integrated Gaussian curvature of a band between two parallels is $-2\pi$ times the jump of $R' = \sin\alpha$ across the band (from $K = -R''/R$ and $d^2\mu = R\,du\,dv$). In the hourglass, nanotube ($R'=0$), nanocone ($R' = \sin\alpha_n = n/6$) and graphene sheet ($R'=1$) are all flat, and the curvature is concentrated on the junctions: $-2\pi\,n/6$, i.e., $n$ units of $-\pi/3$, on each CNT-CNC junction, and $-2\pi(6-n)/6$, i.e., $6-n$ units, on each CNC-sheet junction. The total, $-4\pi$, i.e., twelve units, is exactly that of HYP (Table~\ref{tab:GBbalance}), as it must be, since both connect a tube-like neck to two flat sheets. The difference is in the distribution: in the hourglass the twelve units must be carried by defects concentrated along four circular junctions, that in Ref.~\protect\refcite{hourglass2021} are idealized as unbonded joints between three distinct nanostructures; in HYP they are spread over the whole surface, with $\sin\alpha = (c/r)\sinh(u/r)$ interpolating continuously between the nanotube-like waist and the flat sheets.

This is why there are clear indications \cite{Iorio2024DesalinationPseudospheres} that, if one realizes carbon-made HYPs with the appropriate $r$ and $c$ parameters, one solves two problems at once: the HYP structure is unique, not made of three pieces, so that no junctions of the CNCs with the CNT are necessary; and the two ends of the carbon HYP are flat along the meridian direction, so that they fit smoothly with the planar graphene, see Fig.~\ref{fig:desalination}(b) and Fig.~\ref{fig:intro}. Such a structure, a carbon-made HYP joining two graphene sheets, was realized in silico, for $r=c$, in Ref.~\protect\refcite{klaver2025stability} (Section~\ref{sec:HYPextrusion} and Fig.~\ref{fig:HYPextrusion}), and proved mechanically stable. Moreover, being able to tune the control parameters, $r$ and $c$, according to needs and requirements (the radius of the neck, $c$, and the ratio between mouth and neck, $\sqrt{1+r^2/c^2}$), is potentially interesting for creating prototypes of desalination or purification filters. A caveat is in order. The ordered, shell-like (``plug'') flow of water in nanotubes relies on a constant radius, while in HYP the radius is only approximately constant around the waist ($R \le 1.1\,c$ over a length $\simeq 0.9\,r$). Whether such a neck suffices, and for which values of $r/c$, has to be settled by dedicated simulations of water transport. The final structure should be a matrix of HYPs between planar graphene sheets, as in Fig.~\ref{fig:desalination}(c), see Ref.~\protect\refcite{Iorio2024DesalinationPseudospheres}.

A route to an experimental realization of carbon structures with this geometry is offered by zeolite-templated carbons (ZTCs), obtained by chemical vapor deposition of a carbon precursor inside the channels of a zeolite, followed by the removal of the template. The carbon inherits the three-dimensional topology of the zeolite, and in frameworks such as FAU (faujasite) and BEA (zeolite beta) the intersections of the channels produce hourglass-shaped regions of negative Gaussian curvature, with very narrow distributions of micropores, that can be regarded as experimentally accessible approximants of carbon HYPs. This is the strategy of a research project on the purification and desalination of water, also for lunar applications, that will interpret ZTCs as realizations of carbon pseudospheres \cite{ZeoLunar2026}.

\section{Conclusions and future prospects}
\label{sec:Summary}

We have reviewed and extended the theory and practice of carbon-made Lobachevsky surfaces, organizing them around the balanced Gauss--Bonnet theorem: the total curvature fixed by the topology is shared among disclination defects, elastically deformed hexagons, and boundaries.

On the side of the defect-free curvature, we have shown that elastic deformations can carry intrinsic curvature, but cannot change the topology: a sphere made of hexagons only is a punctured sphere, as shown explicitly by the inverse stereographic image of the honeycomb, whose deficit angles add up to $4\pi$ only in the limit, at the missing point. Defects take over above a threshold set essentially by the integrated curvature, one sixth of the sphere for a pentagon.

On the side of the defects, we have shown that the discrete symmetries of the defects on the three pseudospheres stem from the elliptic (ELL), parabolic (BEL) and hyperbolic (HYP) one-parameter subgroups of $SL(2,\mathbb{R})$, and that the defects sit where the surface has accumulated an integer number of units of curvature, $\pi/3$. We have provided the explicit arrangement for the whole family of ELLs, recovered the group $F_Y$ of BEL as their limit, and proposed the corresponding arrangement for HYP.

On the side of the actual carbon structures, we have confronted ideal HYPs by design with HYPs obtained by simulated extrusion and annealing: the total curvature charge, $-4\pi$, is the same; its distribution is completely different, and the shape is recovered only when the boundary term is controlled.

Two lines of research are open. The first is quantitative: as seen with large fullerenes, and with the large ideal HYPs of Section~\ref{sec:HYPdesign}, the larger the carbon structure with isolated defects, the closer it is to a triangulated polyhedron, the hexagonal regions between the defects tending to be flat. Keeping the curvature between the defects requires deformed hexagons, i.e., stress, that can be imposed by a substrate or by the boundaries, as for the HYP embedded in flat sheets. A theory of how the non-topological curvature of the deformed hexagons and the topological curvature of the defects share the total curvature, for arbitrary configurations, is the main open problem that this work leaves us with.

The second is structural, and it is the one that motivated this program from the start. The finite subgroups of $SO(3)$ are few, and they are tied to the five Platonic solids. For the Lobachevsky geometry, this finiteness breaks down: $SO(2,1)$ has infinitely many discrete subgroups, the Fuchsian groups, among which are the infinitely many hyperbolic triangle groups, symmetry groups of infinitely many ``Lobachevsky Platonic solids''. At the same time, there are infinitely many Lobachevsky surfaces in $\mathbb{R}^3$, each realizing a portion of the Lobachevsky plane. We have seen that the three pseudospheres, the simplest of them, correspond to the three classes of one-parameter subgroups of $SL(2,\mathbb{R})$, with the discrete structures $F_{C_{k-1}}$, $F_Y$ and $F_{D_6}$. It is tempting to conjecture a correspondence between the two infinities, discrete groups on one side and surfaces on the other, of which the pseudospheres are the first instance. To our knowledge, the crystallography of the Lobachevsky plane has been developed for hyperbolic lattices \cite{Maciejko2021HyperbolicBand,Boettcher2022Crystallography} and for periodic minimal surfaces \cite{SadocCharvolin1990,SadocMosseri1999}, but not in connection with the constant-curvature surfaces embedded in $\mathbb{R}^3$. The next steps are then the construction of the full discrete groups for HYP and for the other Lobachevsky surfaces, such as the Dini surface, and the laboratory realization of these structures.

\section*{Acknowledgements}

P.C., R.G., A.I. and P.P. gladly acknowledge support from Charles University Research Center (UNCE 24/SCI/016). T.P.C.K. and D.L. acknowledge support by the Ministry of Education, Youth and Sports of the Czech Republic through the e-INFRA CZ (ID:90254). O.D.L. and R.G.A. acknowledge the STAR funding: Progetto STAR X -- 2583\_QIIR111\_00544, CUP H23C26000120007, PN RIC 2021-2027, Italian Ministry of University and Research. A.I. thanks the VIS programme of the Regione Calabria for a Visiting Professorship at the University of Calabria (UNICAL), where large parts of this work were carried out.

\vspace{0.2in}
\textbf{Declarations} The authors declare no competing interests and no conflicts.

\textbf{Data availability statement} All data that support the findings of this study are included within the article. The scripts used for Section~\ref{sec:StereoHexagons} are available as supplementary material; the input and output files of the simulations of Ref.~\protect\refcite{klaver2025stability} are available with that work; other data are available from the authors upon reasonable request.

\startappendix

\myappendix{Deficit angle and Gaussian curvature of discrete structures}\label{app:DeficitAngles}

The Gaussian curvature $K(u,v)$ is an intrinsic property of a surface, given at any point by the \textit{product} of the two principal curvatures at that point. It opposes to extrinsic, or mean, curvature, $M$, that is (half) the \textit{sum} of the principal curvatures. We shall give other formulae for $K$ later here. For the moment see Ref.~\protect\refcite{Gauss1827} and Ref.~\protect\refcite{Spivak1979}.

With this, surfaces with any (smooth enough) shape in one direction, and flat in the orthogonal direction, are intrinsically flat, $K=0$, even though $M \neq 0$ everywhere.

We shall be dealing with discrete structures, mostly hexagonal lattices. We shall also allow for deformations of the polygons. We then allow the edges to have different lengths, but not to bent, for the moment. One can think of those edges as elastic, but in the plane of the polygon.

\begin{figure}[htbp]
  \centering
  \includegraphics[width=\linewidth]{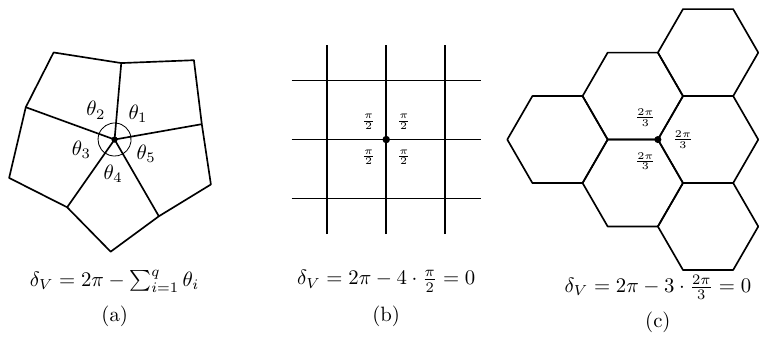}
  \caption{(a) The angles $\theta_i$ of the $q$ polygons meeting at a vertex and the deficit angle $\delta_V$; (b) square and (c) hexagonal tilings of the plane: zero deficit angle at every vertex.}
  \label{fig:vertextilings}
\end{figure}

For such structures the Gaussian/intrinsic curvature is given by the sum of the deficit ($\delta > 0$) or excess ($\delta < 0$) angles at every vertex of the structure
\begin{equation}\label{defangApp}
  \delta = 2 \pi - \sum_{i = 1}^{q} \theta_i \;,
\end{equation}
with $\theta_i$ the angles at the vertex of the $q$, in general irregular, polygons, meeting there. See Fig.~\ref{fig:vertextilings}(a). The case of tessellations of regular polygons is clearly included in this general definition.

Locally the curvature is then given, vertex by vertex, by $K_V = \delta_V$, while, for a compact structure (no boundaries)
\begin{equation}\label{KtotV}
  K_{tot} = \sum_{all \, vertices} K_V \;.
\end{equation}

When there are boundaries, the vertices on the boundary carry the discrete counterpart of the geodesic curvature, that is the exterior angle $\pi - \sum_i \theta_i$, with $\theta_i$ the angles of the polygons meeting there, cf. Eq.~\eqref{GaussBonnetCorners}, and the Gauss--Bonnet theorem reads
\begin{equation}\label{Gauss-Bonnet generic}
  \sum_{V \in \, {\rm bulk}} \Big( 2\pi - \sum_{i} \theta_i \Big) \; + \sum_{V \in \, {\rm boundary}} \Big( \pi - \sum_{i} \theta_i \Big) = 2 \pi \, \chi \;.
\end{equation}
The first sum is $K_{tot}$, the second is the boundary term: the boundaries contribute to $\chi$, but not to the total Gaussian curvature. This is important when counting the number of necessary defects for the pseudospheres, that do have boundaries, see Refs.~\protect\refcite{weylgraphene,LobLab_2016} and Table~\ref{tab:VariousEuler}.

Two elementary checks. In the square, hexagonal and triangular tilings of the plane (Fig.~\ref{fig:vertextilings}(b),(c) and Fig.~\ref{PlaneTessel}) $\delta = 2\pi - 4\cdot\frac{\pi}{2} = 2\pi - 3\cdot\frac{2\pi}{3} = 2\pi - 6\cdot\frac{\pi}{3} = 0$ at every vertex, so that $K_V = 0$ and $K_{tot}=0$. For the tetrahedron, $\delta = 2\pi - 3\cdot\frac{\pi}{3} = \pi$ at each of its 4 vertices, and for the cube $\delta = 2\pi - 3\cdot\frac{\pi}{2} = \frac{\pi}{2}$ at each of its 8 vertices: in both cases $K_{tot} = 4\pi = 2\pi\cdot2$, the correct total curvature for $\chi = 2$.

\myappendix{An illustrative counting}\label{app:CountingHYP}

We report here the analysis of the output of the simulation of the carbon HYP obtained by nano-extrusion and annealing (Section~\ref{sec:HYPextrusion}). We do it to show the details of the counting of the intrinsic curvature associated with the discrete structure, that, although elementary, can be easily overlooked. The structure analyzed is one of the HYPs embedded between the two periodic graphene sheets, and the scars that extend across the Hilbert horizons into the sheets are counted in full. In the units used here the total curvature of the sphere is $1$, so that $-4\pi$ corresponds to $-1$, i.e., twelve heptagons of $-1/12$ each.

The output file was analyzed to characterize morphology and arrangement of the defects. A defect was defined as a non-hexagonal ring, or a set of adjacent ones, and/or non-$sp^2$ carbons.

Non-hexagonal rings were mostly 7- and 5-sided. They formed thread-like structures, often one ring thick or, less commonly, two. These structures, also known as scars, are in most cases linear but they can bifurcate, typically branching at 7- and 8-sided polygons. One atom was found to be $sp$-hybridized (2 bonds) and three others were $sp^3$ (4 bonds).

Heptagonal rings introduce -1/12, octagonal rings -1/6 and so on. In general, n-gons bring in (6-n)/12 of curvature. $sp$-hybridized carbons bring in 1/6 while $sp^3$ -1/6.

\begin{figure}[htbp]
  \centering
  \subfloat[isolated heptagon, $-1/12$]{\includegraphics[width=0.32\linewidth]{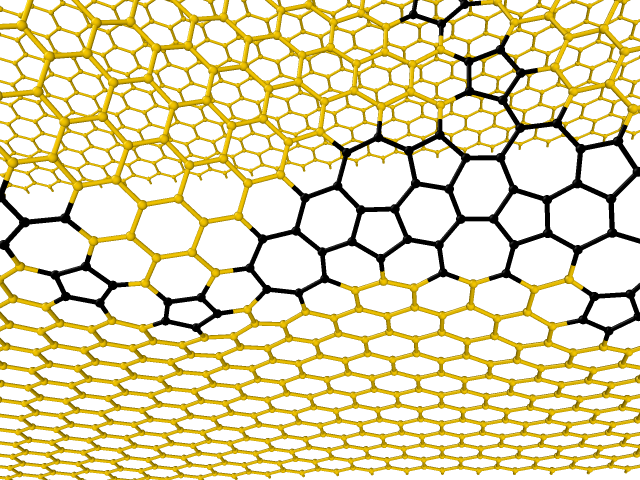}\label{subfig:hept}}\hfill
  \subfloat[scar 2, $-1/12$]{\includegraphics[width=0.32\linewidth]{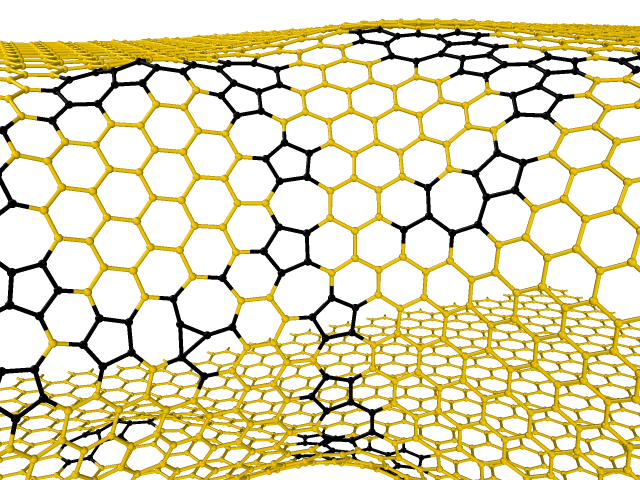}\label{subfig:scar2}}\hfill
  \subfloat[scar 5, $-1/6$]{\includegraphics[width=0.32\linewidth]{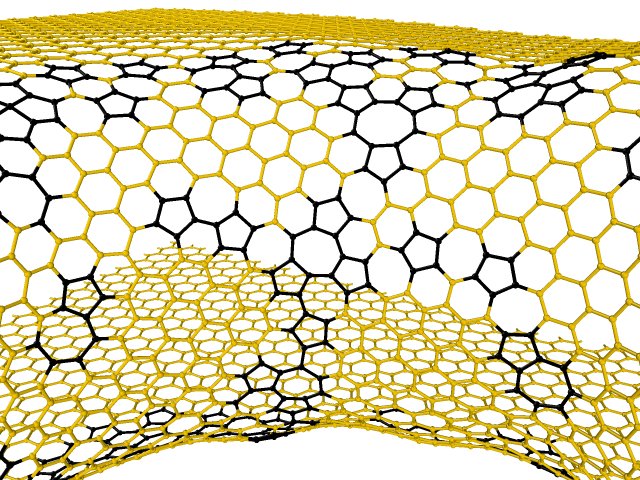}\label{subfig:scar5}}\\
  \subfloat[scar 9, $+1/12$]{\includegraphics[width=0.32\linewidth]{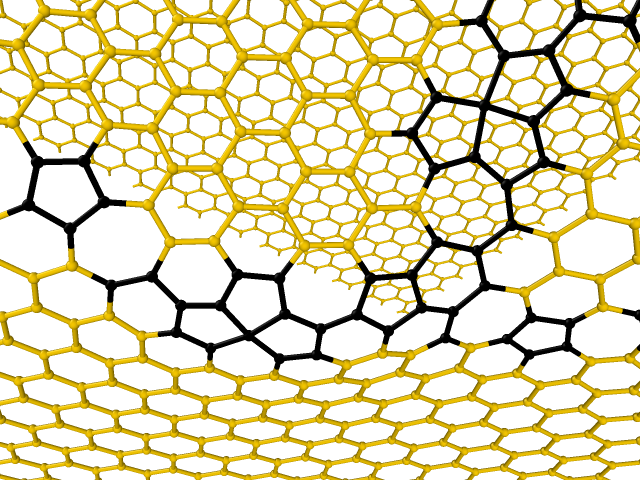}\label{subfig:scar9}}\hfill
  \subfloat[scar 12, $0$]{\includegraphics[width=0.32\linewidth]{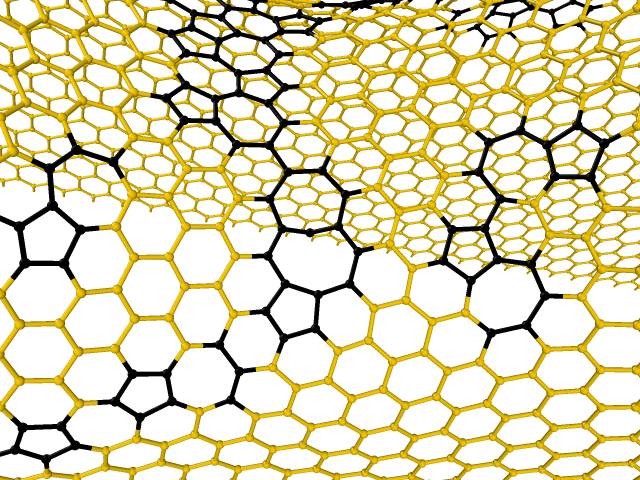}\label{subfig:scar12}}\hfill
  \subfloat[scar 1, $-1/12$]{\includegraphics[width=0.32\linewidth]{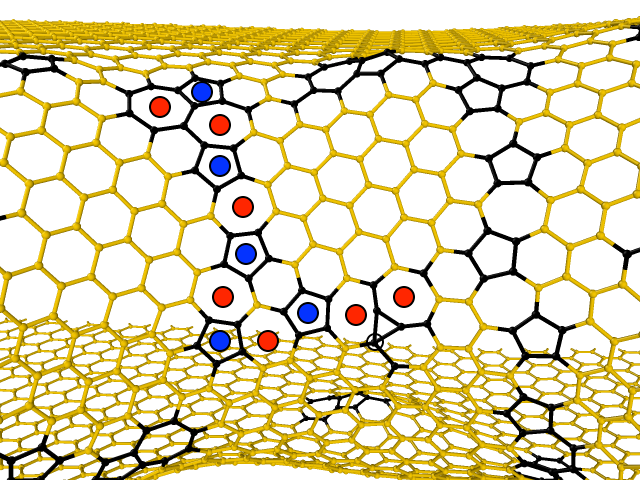}\label{subfig:scar1}}
  \caption{Representative defects of the carbon HYP of Ref.~\protect\refcite{klaver2025stability} (see Table~\ref{tab:scars}), with their curvature in units of $4\pi$. Scar 9 (d) contains an $sp^3$ carbon, scar 12 (e) an $sp$ carbon, and scar 1 (f), on top of two extra heptagons ($-1/6$), a triangular ring ($+1/4$) with an $sp^3$ carbon ($-1/6$), so that it carries $-1/12$ overall.}
  \label{fig:scars}
\end{figure}

Visual inspection identified the entire expected total curvature. The fifteen defects found are listed in Table~\ref{tab:scars}, and six representative ones are shown in Fig.~\ref{fig:scars}. All pictures have been generated with OVITO (Open Visualization Tool) \cite{Stukowski:2010aa}, Basic 3.11.0, whose structure identification tool can match various templates, including graphene. For larger structures, an automated method, such as those implemented in polymake \cite{Gawrilow:2000,Assarf:2017aa} or rdkit, would be recommended.

\begin{table}[htbp]
\centering
\tbl{Charged and most extensive neutral defects of the carbon HYP obtained by nano-extrusion and annealing \cite{klaver2025stability}, with their curvature in units of $4\pi$ (one heptagon gives $-1/12$). Scar 4 is split in two parts, one above and one below. The charge of a scar is assigned, conventionally, to one of its ends, and some scars cross from one half to the other: only the total is significant, not the partial sums of the two halves.\label{tab:scars}}
{\begin{tabular}{|c|c|l|c|}
    \hline
    half & defect & notes & curvature \\ \hline
    top    & scar 1          & triangular ring with an $sp^3$ carbon & $-1/12$ \\
    top    & scars 2, 3      & & $-1/12$ each \\
    top    & scar 4 (above)  & neutral & $0$ \\
    top    & scar 4 (below)  & & $-1/12$ \\
    top    & scar 5          & & $-1/6$ \\
    top    & scars 6, 7      & & $-1/12$ each \\ \hline
    bottom & scar 8          & & $-1/12$ \\
    bottom & scar 9          & contains an $sp^3$ carbon & $+1/12$ \\
    bottom & scar 10         & contains an $sp^3$ carbon & $-1/12$ \\
    bottom & scar 11         & & $-1/12$ \\
    bottom & scar 12         & contains an $sp$ carbon, neutral & $0$ \\
    bottom & scar 13         & & $-1/12$ \\
    bottom & isolated heptagon & & $-1/12$ \\ \hline
    \multicolumn{3}{|r|}{total} & $-1$, i.e., $-4\pi$ \\ \hline
  \end{tabular}}
\end{table}

\end{document}